\documentclass[aps,prd,final,reprint,showkeys,showpacs,superscriptaddress,floatfix,nofootinbib,twocolumn,amsmath,amssymb]{revtex4-1}
\usepackage[utf8]{inputenc}

\usepackage{hyperref} 
\usepackage{url}
\usepackage{graphicx}

\usepackage{graphics}
\usepackage{natbib}
\graphicspath{{./Figure}}

\usepackage[normalem]{ulem}
\usepackage{color}

\usepackage{soul}

\def\units#1{~\hbox{$\,{\rm #1}$}}
\def\degrees{\hbox{$^\circ$}}

\begin{document}

\title{A search for dark matter  cosmic-ray electrons and positrons from the Sun with the Fermi Large Area Telescope }

\author{A.~Cuoco}
\email{cuoco@physik.rwth-aachen.de}
\affiliation{Istituto Nazionale di Fisica Nucleare, Sezione di Torino, via Pietro Giuria 1,  I-10125 Torino, Italy}
\affiliation{Dipartimento di Fisica, Università di Torino, Via P. Giuria 1, 10125 Torino, Italy}
\affiliation{RWTH Aachen University, Institute for Theoretical Particle Physics and Cosmology (TTK), D-52056 Aachen, Germany}
\affiliation{Univ. Grenoble Alpes, USMB, CNRS - Laboratoire d'Annecy de Physique des Particules, 9 Chemin de Bellevue, F-74940 Annecy, France}
\author{P.~De~La~Torre~Luque}
\affiliation{Istituto Nazionale di Fisica Nucleare, Sezione di Bari, via Orabona 4, I-70126 Bari, Italy}
\affiliation{Dipartimento di Fisica ``M. Merlin" dell'Universit\`a e del Politecnico di Bari, via Amendola 173, I-70126 Bari, Italy}
\author{F.~Gargano}
\affiliation{Istituto Nazionale di Fisica Nucleare, Sezione di Bari, via Orabona 4, I-70126 Bari, Italy}
\author{M.~Gustafsson}
\affiliation{Georg-August University G\"ottingen, Institute for theoretical Physics - Faculty of Physics, Friedrich-Hund-Platz 1, D-37077 G\"ottingen, Germany}
\author{F.~Loparco}
\email{francesco.loparco@ba.infn.it}
\affiliation{Istituto Nazionale di Fisica Nucleare, Sezione di Bari, via Orabona 4, I-70126 Bari, Italy}
\affiliation{Dipartimento di Fisica ``M. Merlin" dell'Universit\`a e del Politecnico di Bari, via Amendola 173, I-70126 Bari, Italy}
\author{M.~N.~Mazziotta}
\email{mazziotta@ba.infn.it}
\affiliation{Istituto Nazionale di Fisica Nucleare, Sezione di Bari, via Orabona 4, I-70126 Bari, Italy}
\author{D.~Serini}
\affiliation{Istituto Nazionale di Fisica Nucleare, Sezione di Bari, via Orabona 4, I-70126 Bari, Italy}
\affiliation{Dipartimento di Fisica ``M. Merlin" dell'Universit\`a e del Politecnico di Bari, via Amendola 173, I-70126 Bari, Italy}

\date{\today}

\begin{abstract}
We use 7 years of electron and positron Fermi-LAT data to search for a possible excess in the direction of the Sun in the energy range from 42 GeV to 2 TeV. In the absence of a positive signal we derive flux upper limits which we use to constrain two different dark matter (DM) models producing $e^+ e^-$ fluxes from the Sun. In the first case we consider  DM model being captured by the Sun due to elastic scattering and annihilation into $e^+ e^-$ pairs via a long-lived light mediator that can escape the Sun. In the second case we consider instead a model where DM density is enhanced around the Sun through inelastic scattering and the DM annihilates directly into $e^+ e^-$ pairs. In both cases we perform an optimal analysis, searching specifically for the energy spectrum expected in each case, i.e., a box-like shaped and line-like shaped spectrum respectively. 
No significant signal is found and we can place limits on the spin-independent cross-section in the range from $10^{-46}\units{cm^2}$ to $10^{-44}\units{cm^2}$ and on the spin-dependent cross-section in the range from $10^{-43}\units{cm^2}$ to $10^{-41}\units{cm^2}$. 
In the case of inelastic scattering the limits on the cross-section are in the range from $10^{-43}\units{cm^2}$ to $10^{-41}\units{cm^2}$. The limits depend on the life time of the mediator (elastic case) and on the mass splitting value (inelastic case), as well as on the assumptions made for  the size of the deflections of electrons and positrons in the interplanetary magnetic field.
\end{abstract}

\pacs{95.35.+d, 95.85.Ry}

\keywords{Cosmic-ray Electrons and Positrons, Dark Matter}

\maketitle

\section{Introduction}
\label{sec:intro}
The Sun is a promising target for indirect searches for dark matter (DM). DM particles from the Galactic halo passing through the Sun are expected to elastically scatter off nuclei, lose energy, and be trapped by the solar gravitational field. Multiple scatterings cause DM to accumulate in the centre of the Sun. If DM particles can self-annihilate, the capture is then balanced by the annihilation leading to an equilibrium situation. 

Neutrinos produced in the annihilation can then be detected at Earth and can therefore be a probe of the DM scattering cross section on nucleons. Other standard Model (SM) particles produced in the annihilation will be, instead, absorbed in the Sun before being able to escape. However, alternative models, in which DM annihilates into a long-lived mediator that can escape and decay outside the Sun into SM particles, have also been proposed~\cite{Pospelov:2007mp,ArkaniHamed:2008qn,Schuster:2009fc,Leane:2017vag,Schuster:2009au,Bell:2011sn,Arina:2017sng}. In the framework of these models, gamma rays and electrons produced from DM captured by the Sun are also detectable at Earth. In a further class of models, DM scatters inelastically off Sun nuclei, instead of elastically (see for instance~\cite{TuckerSmith:2001hy,Finkbeiner:2009ug,Chang:2008gd,Menon:2009qj,Nussinov:2009ft,Catena:2018vzc,Blennow:2018xwu}). This leads to a significant accumulation of DM particles just outside the surface of the Sun. Within these models, then, gamma rays and electrons are detectable also in the case of direct DM annihilation without a mediator.

In Ref.~\cite{Ajello:2011dq} a search for cosmic-ray electrons and positrons (CREs) from the Sun was performed using data Large Area Telescope (LAT)~\cite{Atwood:2009ez}  on board the Fermi Gamma-ray Space Telescope (FGST)\footnote{The LAT cannot distinguish between electrons and positrons and we refer to both as cosmic ray electrons (CREs).}. In the present work, we update the previous analysis using CREs collected in almost 7 years by the LAT and analyzed with the newest Pass 8 event selection~ \cite{Abdollahi:2017nat}, in a broader energy range from 42\units{GeV} to 2\units{TeV} with an improved analysis method. 

The LAT observes the entire sky every 2~orbits ($\sim$3 hours) when the satellite is operated in the usual ``sky-survey mode'', keeping the Sun in its field of view for a large amount of its live time. The Sun is not expected to be a source of high-energy electrons and positrons. Nevertheless, the interactions of cosmic rays (mainly protons) with its surface yield secondary particles in the hadronic cascades, that can escape from the Sun, such as high-energy gamma rays or neutrinos~\cite{Orlando:2006zs,Orlando:2008uk,Abdo:2011xn,Tang:2018wqp,Seckel:1991ffa,Edsjo:2017kjk}. Soft secondary charged particles are also produced in these interactions, but most of them are expected to be trapped in the Sun due to the intense magnetic field in the corona and in the photosphere. The magnetic field in the region of the solar surface can in fact reach values above a few Gauss~\cite{hmiweb}, resulting in Larmor radii smaller than the solar radius for CREs with energies below $10\units{TeV}$. Those CREs produced in hadronic interactions and eventually escaping from the Sun should exhibit a featureless energy spectrum.

The solar magnetic field also affects the propagation of those charged particles which are not trapped in the Sun. Therefore, CREs produced in the Sun (or in the region between the Sun and the Earth's orbit) and reaching the Earth will not travel along straight paths, but their trajectories will be curved, and their arrival directions at Earth will not point back to the Sun. In addition, since the solar magnetic field is strongly varying with time, the trajectories followed by solar CREs arriving at Earth will also change with time. As a consequence, we expect that the arrival directions of solar CREs at Earth will be spread over a cone with its axis pointing to the Sun and with an aperture that accounts for these effects. The implementation of a detailed study of the propagation of CREs, which takes into account the complex spatial structure of the solar magnetic field and its time evolution, is not trivial and goes beyond the goals of the present work. Therefore, in the calculations of section~\ref{sec:dmmodels} we will make the simplifying assumption that CREs originating from DM annihilations in the region between the Sun and the Earth will reach the Earth after propagating along straight lines.

In the previous work~\cite{Ajello:2011dq} the flux asymmetry towards the Sun and the anti-Sun directions was used to constrain the possible presence of Sun CREs. No significant excess from the Sun was observed, and the upper limits on the excess flux were used to constrain the DM-nucleon scattering cross sections for two DM models. In the first, DM annihilates in the core of the Sun into two long-lived mediators that decay outside the Sun into two electron-positron pairs, while in the second DM annihilates outside the Sun directly into electron-positron pairs. 

In the present work we improve this strategy exploiting the fact that in the two models above, CREs are expected to be produced with a peculiar energy spectrum. We thus search for a signal with a specific spectral shape. In particular, for the case of annihilation through a mediator, we search for a box-like spectrum, which is expected in this scenario~\cite{Ibarra:2012dw}. On the other hand, for the case of direct annihilation into CREs, we search for a line-like monochromatic feature. The method adopted here is based on our recent work~\cite{Mazziotta:2017ruy}, in which we search for features in the energy spectrum of Galactic CREs. 

For the present analysis we have also revised the model describing the production of SM particles from DM annihilations via long-lived mediator, which will be discussed in section~\ref{sec:dmmodels}. In addition, we have also implemented an approach which takes systematic uncertainties into account. The details of the analysis method are discussed in section~\ref{sec:anamet}. In section~\ref{sec:res} we will discuss the results and the differences with respect to our previous work~\cite{Ajello:2011dq}. Finally, in section~\ref{sec:con} we will compare our results with those obtained from similar searches performed with gamma rays.

\section{Dark Matter models from the Sun and electron-positron energy spectrum}
\label{sec:dmmodels}
As mentioned above, in this work we assume two scenarios for the production of CREs due to annihilations of DM particles captured inside the Sun: 
i) capture via elastic scattering and subsequent annihilation in the core of the Sun into $e^+e^-$ pairs through a light intermediate state $\phi$;
ii) capture via inelastic scattering and subsequent annihilation directly into $e^+e^-$ pairs outside the Sun.

\subsection{Annihilation through a light intermediate state}
\label{sec:inter}
We assume a standard scenario, in which DM particles are captured by the Sun through elastic scattering interactions and then continue to lose energy through subsequent scatterings, eventually thermalizing and sinking to the core where they annihilate.
The rate of change of the number of DM particles $N_\chi$ in the Sun at a given time $t$ can be written as a function of the capture and annihilation rates\footnote{We consider DM masses above a few\units{GeV} since for lower masses the particle evaporation can be non-negligible~\cite{Griest:1986yu,Gaisser:1986ha}.} as:

\begin{equation}
\frac{dN_{\chi}}{dt} = \Gamma_{\text{cap}} – C_{\text{ann}} N_\chi^2 
\label{eq:balance}
\end{equation}
where $\Gamma_{\text{cap}}$ is the capture rate and $C_{\text{ann}}$ is a factor accounting for the annihilation cross section. 

When capture and annihilation reach equilibrium ($dN_{\chi}/dt = 0$), the annihilation rate $\Gamma_{\text{ann}} = \frac{1}{2} C_{\text{ann}}N_\chi^2$ is given by:

\begin{equation}
 \Gamma_{\text{ann}} = \frac{1}{2} \Gamma_{\text{cap}}  
 \label{eq:gammaann}
\end{equation}

The factor $1/2$ accounts for the two DM particles involved in each annihilation event. The annihilation rate at equilibrium is independent of the velocity-averaged annihilation cross section $\langle \sigma v \rangle$, and is set by $\Gamma_{\text{cap}}$, which depends on the scattering cross section (spin independent or spin dependent), the local halo DM number density, the DM mass, the DM velocity and its dispersion, i.e., $\Gamma_{\text{ann}} = \Gamma_{\text{ann}} (m_\chi, \sigma, ...)$. 

If equilibrium between capture and annihilation is not reached, the annihilation rate resulting from Eq.~\ref{eq:balance} at any given time $t$ is $\Gamma_{ann}=\frac{1}{2} \Gamma_{\text{cap}} \tanh^{2}(t/\tau)$, where the time scale $\tau=(\Gamma_{cap}C_{ann})^{-1/2}$~\cite{Jungman:1995df}. Therefore the right-hand side of Eq.~\ref{eq:gammaann} should be corrected with the factor $\tanh^{2}(t/\tau)$. In the following we will assume equilibrium between capture and annihilation.

We assume a model in which DM particles annihilate into a light intermediate state $\phi$, i.e., $\chi \chi \rightarrow \phi \phi$, with the $\phi$ subsequently decaying to standard model particles. The $\phi$ are assumed to be able to escape the Sun without further interactions, with each $\phi$ decaying to an $e^\pm$ pair. If this decay happens outside the surface of the Sun, the $e^\pm$ can reach the Earth and may be detectable in the form of an excess of CREs from the direction of the Sun.

The DM particles are assumed to annihilate at rest in the core of the Sun. Therefore, in the lab frame, the energy of the $\phi$ will be equal to the mass of the DM particle, i.e. $E_\phi = m_\chi$. We assume $\phi$ to be a light scalar such that $m_\phi \ll m_\chi$. Under this assumption, the $\phi$ are relativistic, i.e. $\gamma_\phi = E_\phi/ m_\phi \gg 1$.
The angular dispersion of the $e^\pm$ pair with respect to the direction of the parent $\phi$ is of the order of $1/\gamma_{\phi} \ll 1$, and therefore the $e^\pm$ will keep the same direction as the $\phi$. As a consequence, the flux of $e^\pm$ is equivalent to that from a point-like source centered in the core of the Sun. 

Indicating with $L$ the $\phi$ decay length, the flux of $e^{\pm}$ produced from $\phi$s decaying at a distance $r$ from the center of the Sun and detected at the Earth is given by:

\begin{equation}
\frac{d\Phi_{\text{DM}} (E)}{dr} = N(E) \frac{\Gamma_{\text{cap}}}{4 \pi r^2}  \frac{e^{-r/L}}{L}  \frac{r^2}{D^2}
\label{eq:eq3}
\end{equation}
where $\Gamma_{\text{cap}}/4\pi r^2$ is the flux of $\phi$s at the distance $r$ from the Sun, $e^{-r/L}/L$ is the probability density that a $\phi$ decays at the distance $r$, the factor $r^2/D^2$ accounts for the ratio between the surface of the sphere of radius $r$ and the surface of the sphere or radius corresponding to the Sun-Earth distance $D$, and $N(E)$ is the DM $e^\pm$  spectrum per $\phi$ decay. The $e^\pm$ flux at the Earth is calculated integrating Eq.~\ref{eq:eq3} from the Sun ($r=R_\odot$) to the Earth ($r=D$) and is given by:

\begin{equation}
\Phi_{\text{DM}}(E) =  N(E) \frac{\Gamma_{\text{cap}}}{4 \pi D^2} \left( e^{-R_\odot/L} - e^{-D/L} \right) 
\label{eq:phidm}
\end{equation}

The $e^\pm$ flux at the Earth due to the DM annihilations on the Sun via long-lived mediator is then similar to the flux from a point-like source in the Sun ($\frac{\Gamma_{\text{cap}}}{4 \pi D^2}$) modulated by the survival probability of the mediator ($e^{-R_\odot/L} - e^{-D/L}$).  The DM $e^\pm$ spectrum in this scenario acquires a box-like shape centered on $E=m_\chi/2$, with a width depending on the mediator mass $m_\phi$~\cite{Ibarra:2012dw}. As mentioned above, we are assuming $m_\phi \ll m_\chi$. In this limit the dependence on $m_\phi$ disappears and the box extends from $E=0$ to $E=m_\chi$. We can therefore write $N(E)=2 H (m_\chi-E) /m_\chi$, with the factor $2$ accounting for the $e^\pm$ multiplicity ($2$) for each mediator and $H$ the Heaviside step function. We assume that mediators can pass through Sun without attenuation.
\footnote{Strictly speaking, the spectrum is box-shaped only when the whole angular extent of the emission is integrated. The emission is, indeed, not exactly point-like and the spectrum depends on the observing direction angle with respect to the line-of-sight, and becomes box-shaped when summed over all the directions~\cite{Ibarra:2012dw}. On the other hand, if only $e^{\pm}$ at small angles with respect to the line-of-sight are considered, deviations from the box shape  occur at low energies. Since the angular dispersion of the $e^{\pm}$ pair with respect to the direction of the $\phi$ is $\sim 1/\gamma$, in the limit we consider, i.e., $m_\phi \ll m_\chi$, the extent of the emission is much smaller than size of the region of interest we use in the analysis. Hence we are effectively integrating all of the emission, so that the box-shaped spectrum is a valid assumption.}

\begin{figure}[!ht]
\centering
\includegraphics[width=\columnwidth,height=0.25\textheight,clip]{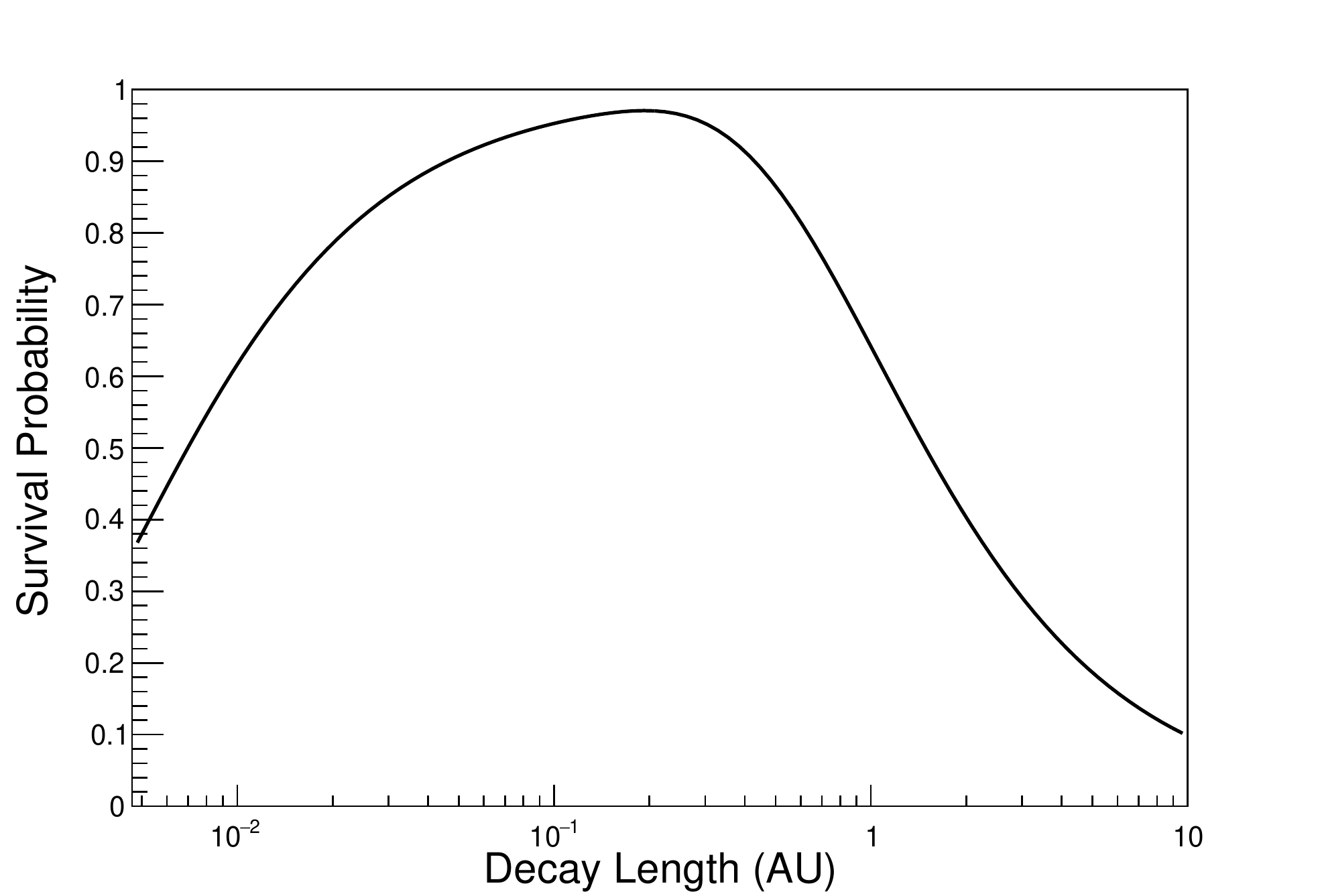}
\caption{Probability that CREs produced in the decays of a mediator travelling from the Sun to the Earth can reach the Earth as a function of the decay length of the mediator.}
\label{fig:fig1}
\end{figure}

Fig.~\ref{fig:fig1} shows the survival probability as a function of the mediator decay length $L$, which in turn depends on the life time $\tau$ as $L=\gamma c \tau$. The CRE signal is possible only if the mediator decays outside the Sun and before the Earth. The survival probability shows a maximum for a decay length of about $0.3\units{AU}$ and its value changes of about a factor $2$ for decay lengths in the range $R_{\odot} - D$.

The capture rate has been evaluated with the {\tt DARKSUSY} code version {\tt 6.1.0}~\cite{Gondolo:2004sc,Bringmann:2018lay,darksusyweb} assuming the default settings, with a local DM density $\rho_\odot = 0.3\units{ GeV/cm^3}$, a Maxwellian velocity distribution with average velocity $v_\odot=220\units{km/s}$ and velocity dispersion $v_{rms}=270\units{km/s}$, and setting the DM-nucleon cross section to $\sigma=10^{-40}\units{cm^2}$ (in both the spin independent and spin dependent cases). 

\begin{figure}[!ht]
\centering
\includegraphics[width=\columnwidth,height=0.25\textheight,clip]{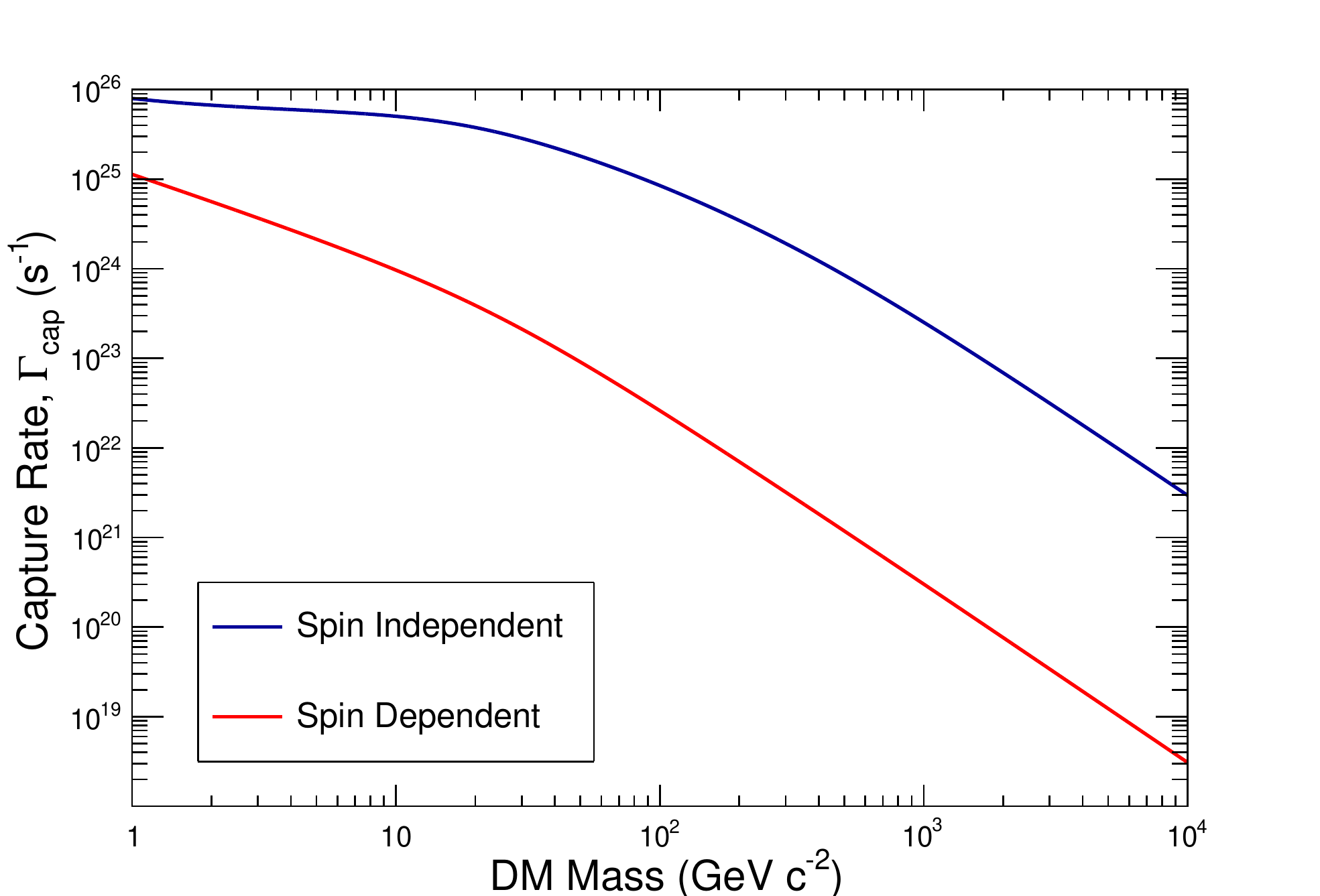}
\caption{Capture rate as a function of the DM mass, for the spin independent (blue line) and spin dependent (red line) cases. In both cases a scattering cross section of $10^{-40}\units{cm^{2}}$ is assumed.}
\label{fig:fig2}
\end{figure}

Fig.~\ref{fig:fig2} shows the capture rate as a function of the DM mass, for the spin independent (blue line) and spin dependent (red line) cases.
The capture rate scales as $m_\chi^{-1}$ up to a few tens of$\units{GeV/c^2}$, following the local DM number density, while above a few hundreds$\units{GeV/c^2}$ it scales as $m_\chi^{-2}$, due to kinematic suppression of the energy loss~\cite{Jungman:1995df,Rott:2012qb}.

\subsection{Inelastic dark matter}
\label{sec:idm}
In the above standard scenario, DM particles undergo subsequent scatterings which cause them to sink to the core of the Sun, and therefore the fraction of captured DM particles outside the surface of the Sun at any given time is negligible~\cite{Sivertsson:2009nx}.

On the other hand, DM  can also scatter inelastically. In this case, a DM particle $\chi$ that scatters inelastically off a nucleon $N$ will produce an excited state $\chi^{*}$ with a slightly heavier mass, i.e. $\chi + N \rightarrow \chi^{*} + N$. To undergo a scattering, the DM particle needs an energy $E \ge \Delta(1+m_\chi/m_N)$, where $\Delta = m_\chi^{*} - m_\chi$ is the mass splitting parameter. In this situation, particles captured by the Sun by inelastic scattering are typically able to undergo only few interactions until their kinetic energy goes below the threshold for further scatterings. If the elastic scattering cross section is sufficiently small ($\sigma<10^{-47}\units{cm^2}$), the captured particles will be unable to thermalize and settle to the core, and instead will remain on relatively large orbits. As a result, the density of captured DM particles outside the Sun may not be negligible, and the annihilation of those particles to $e^\pm$ could thus produce an observable flux of CREs from the direction of the Sun. 

In the following we will assume that DM  particles annihilate at rest and thus the energy of the $e^\pm$ produced in annihilation is given by $E_{e^\pm}=m_\chi$. We will also assume that CREs do not suffer significant energy losses between the production at the surface of the Sun and the arrival at the detector at Earth. Under these assumptions, in this scenario we expect a flux of mono-energetic CREs, i.e. a line-like feature. We will also assume that all annihilations occur close to the surface of the Sun, since the density of DM rapidly decreases with the distance from the Sun~\cite{Schuster:2009fc}. Also in this case, $e^+e^-$ produced inside the surface of the Sun can not escape from the Sun, and thus do not produce a detectable flux at Earth.

In~\cite{Nussinov:2009ft} it is argued that equilibrium between capture and annihilation of DM particles in this scenario  is eventually reached
also for the case of pure inelastic scattering (but see also \cite{Blennow:2018xwu}).
As for the previous case of elastic scattering, we thus
 assume  equilibrium, i.e. $\Gamma_{\text{ann}} = \frac{1}{2} \Gamma_{\text{cap}}$.
Defining $f_{\text{out}}$ as the fraction of DM annihilations occuring outside the Sun at a given time, we can write the following equation:

\begin{equation}
    \Gamma_{\text{ann,out}} = f_{\text{out}} \Gamma_{\text{ann}} = \frac{1}{2} f_{\text{out}} \Gamma_{\text{cap}}
    \label{eq:eq5}
\end{equation}

We use the capture rate $\Gamma_{\text{cap}}$ as a function of DM mass $m_\chi$ and of mass splitting $\Delta$ as given in Fig.~2 of Ref.~\cite{Menon:2009qj}, interpolating the results shown in that figure for DM masses below $1\units{TeV}$ and extrapolating for masses above this value. These capture rates were calculated assuming a local DM density $\rho_\odot = 0.3\units{ GeV/cm^3}$, a Maxwellian velocity distribution with average velocity $v_\odot=250\units{km/s}$ and velocity dispersion $v_{rms}=250\units{km/s}$, and by setting $\sigma_0=10^{-40} \units{cm^2}$, which the authors of  Ref.~\cite{Menon:2009qj} define as the inelastic  DM-nucleon cross section in the limit $\Delta \rightarrow 0$. The capture rate scales linearly with $\rho_\odot$ and $\sigma_0$, while the dependence on $v_\odot$ and $v_{rms}$ is mild in the mass range considered in the present work. The constraints calculated by direct detection experiments are more sensitive to the velocity distribution of DM in the solar system.

We use the parameter $f_{\text{out}}$ as a function of the splitting mass parameter $\Delta$ as calculated by Ref.~\cite{Schuster:2009fc} for the DM captured by the Sun via inelastic scattering and $m_\chi=1\units{TeV}$. We assume that the dependence on $m_\chi$ is weak for the mass range of the present work, also following the prescriptions of Ref.~\cite{Schuster:2009fc}. We note that the uncertainties in the calculation of $f_{\text{out}}$ are large, and a detailed study is beyond the scope of this paper. However, the present results can be easily rescaled. 

The isotropic flux of $e^\pm$ at the Earth due to the DM annihilations captured nearby the Sun via inelastic scattering is

\begin{equation}
    \Phi_{DM}(E) = 2 \frac{\Gamma_{ann,out}}{4 \pi D^2}~\delta (E - m_\chi)
    \label{eq:idm}
\end{equation}
where the factor 2 accounts for the fact that 2 CREs are emitted per annihilation of each pair of DM particles. However, we assume that only one particle of the $e^+e^-$ pair emitted per annihilation nearby the Sun surface can reach the Earth, so we assume that the flux of CRE observed is a factor 2 smaller than that given by Eq.~\ref{eq:idm}, and this would result in conservative limits. In fact, if DM annihilations take place close to the solar surface, the electron and positron will travel in opposite directions and the particle travelling towards the Sun will likely be absorbed. 

\section{Analysis method}
\label{sec:anamet}
For the present work we analyzed the same dataset used in Refs.~\cite{Abdollahi:2017nat} and~\cite{Mazziotta:2017ruy}. Full details on the event selection are given in Ref.~\cite{Abdollahi:2017nat}.
As discussed above, in the present analysis we search for possible local excesses in the count spectra of CREs from a region of interest (RoI) towards the Sun direction. We assume that these excesses can be modeled either with a delta-like or a box-like feature in the CRE energy spectra. We also use, as control region, a RoI towards the anti-Sun direction with same size as the one used for the signal search. The analysis has been performed in RoIs of different angular sizes.

Once folded with the energy response of the LAT, a local delta-like or box-like feature in the CRE spectrum will show up as a broad peak or a smooth edge in the count spectrum, with the same width as the energy resolution of the instrument. Following the approach of Ref.~\cite{Mazziotta:2017ruy}, we have implemented a fitting procedure in sliding energy windows to search for possible local features on the top of a smooth CRE spectrum. 

In each energy window, we model the CRE intensity as $I(E) = I_{0}(E) + I_{f}(E)$, where $I_{0}(E)$ is the smooth part of the spectrum and $I_{f}(E)$ describes the possible feature. Since the energy windows are narrow, we assume that the smooth part of the spectrum can be described by a power-law (PL) model, $I_{0}(E) = k (E/E_{0})^{-\alpha}$, where $\alpha$ is the PL spectral index and the prefactor $k$ (in units of $\units{GeV^{-1}~ m^{-2}~ s^{-1}~sr^{-1}}$) corresponds the CRE intensity at the scale energy $E_{0}$ fixed to $1 \units{GeV}$.
In the present analysis we assume two models for $I_{f}(E)$: (i) a delta-like (hereafter line model) feature, $I_{f}(E) = s ~ \delta (E_{w}-E)$, where $s$ represents the intensity of the line in units of $\units{m^{-2}~s^{-1}~sr^{-1}}$; (ii) a box-like (hereafter box model) feature, $I_{f}(E) = s ~ H(E_{w}-E)$; where $s$ represents the intensity of the box in units of $\units{GeV^{-1}~ m^{-2}~s^{-1}~ sr^{-1}}$. Here we indicate with $\delta$ the  Dirac delta function and with $H$ the Heaviside step function, while $E_w$ is the energy corresponding to the center of the sliding window.

Starting from the model, we can calculate the expected counts in the $j$-th bin of observed energy in the signal region towards the Sun (S) and in the control region towards the anti-Sun (A) as:

\begin{align}
\label{eq:expcounts}
\mu^{S}_j & = t \int dE ~ \mathcal{R}^{S}(E_j | E) ~ I^{S}(E) \notag \\
 & \\
\mu^{A}_j & = t \int dE ~ \mathcal{R}^{A}(E_j | E) ~ I^{A}(E) \notag
\end{align}
where $E_j$ is the observed energy, $E$ is true (Monte Carlo) energy, $\mathcal{R}(E_j | E)$ is the instrument response matrix (acceptance in units of $\units{m^{2}~ sr}$), which incorporates the energy resolution of the LAT, and $t$ is the integrated livetime.

In our analysis we fit at the same time both the Sun and the anti-Sun count distributions. 
In the null hypothesis we have no signal from  both the Sun and the Anti-Sun and thus we assume that the count distributions can be fitted with the same PL, i.e. $I^{S}(E) = I^{A}(E) = k  (E/E_{0})^{-\alpha}$. On the other hand, in the alternative hypothesis of a CRE signal from the Sun, we assume that the Sun count distribution can be fitted with the sum of a PL and a feature, i.e. $I^{S}(E) = k (E/E_{0})^{-\alpha}+I_{f}(E)$, while the Anti-Sun count distribution is again fitted with the same PL as the Sun, i.e. $I^{A}(E) = k (E/E_{0})^{-\alpha}$.

In our fits we minimize a $\chi^2$ function defined as~\cite{Mazziotta:2017ruy}:

\begin{equation}
  \chi^2 = \sum_{j=1}^{N} \left( \frac{ \left( n^S_j - \mu^S_{j} \right) ^2}{ n^S_j + \left( f_{syst} n^S_j \right)^2} + \frac{ \left( n^A_j - \mu^A_{j} \right) ^2}{ n^A_j + \left( f_{syst} n^A_j \right)^2} \right)
   \label{eq:likelihood}
\end{equation}
where $N$ is the number of energy bins used for the fit, $n^{S}_{j}$ and $n^{A}_{j}$ are the counts in the $j$-th observed energy bin from the Sun and from the Anti-Sun respectively and $\mu^{S}_{j}$ and $\mu^{A}_{j}$ are the corresponding expected counts, evaluated from eq.~\ref{eq:expcounts}. The denominator of each term in the summation includes the sum in quadrature of the statistical Poisson fluctuations ($\sqrt{n_j}$) and of the systematic uncertainties ($f_{\text{syst}} n_j$), which are discussed more in detail below.

To estimate the parameters $\{ k, \alpha, s\}$ which minimize the $\chi^{2}$ we use the {\tt MINUIT} code within the ROOT toolkit~\cite{Brun:1997pa,rootweb}; the values of the parameters at a $95\%$ confidence limit (CL) are evaluated using {\tt MINOS} and setting the error confidence level to $2.71$. 

We have performed our fits scanning an energy range extending from $42 \units{GeV}$ to about $2\units{TeV}$. This interval has been divided in $64$ bins per decade, equally spaced on a logarithmic scale. We perform the fit in sliding energy windows with a half-width of $0.35 E_{w}$, that well contain the possible features. In fact, the LAT energy resolution for the CRE selection at $95\%$ containment ranges from about $15\%$ at $42 \units{GeV}$ to about $20\%$ at $1 \units{TeV}$ and increases up to $35\%$ at $2 \units{TeV}$, as shown in Ref.~\cite{Abdollahi:2017nat}. We also tested different energy binnings and  different window sizes yielding comparable results (see~\cite{Mazziotta:2017ruy}).

To account for possible systematic uncertainties originated from the instrument and from the reconstruction procedure~\cite{Abdollahi:2017nat} that might mimic false local features or might mask true ones, we have implemented a data-driven procedure, using only the data from the control region and following a similar approach to that of Ref.~\cite{Mazziotta:2017ruy}. In each energy window we fit the data from the Anti-Sun region with a simple PL model, considering statistical uncertainties only. We then evaluate the fractional residuals $f_{j}=(n^{A}_{j}-\mu^{A}_{j})/\mu^{A}_{j}$ and we calculate the root mean square (RMS) of their distribution. The RMS on the distribution of fractional residuals can be expressed as $f_{\text{RMS}}^{2}=f_{\text{stat}}^{2}+f_{\text{syst}}^{2}$, where $f_{\text{stat}}$ and $f_{\text{syst}}$ are the contributions from statistical and systematic uncertainties respectively. We finally evaluate $f^{2}_{\text{syst}}$ as the difference between the observed RMS and its expected value when only statistical uncertainties are considered. The systematic uncertainties evaluated with this procedure are then implemented in eq.~\ref{eq:likelihood}.

For each energy window we evaluate the local significance of a possible feature considering the $\chi^2$ difference between the alternative hypothesis (line or box signal) and the null hypothesis (PL model) as Test Statistics, i.e. $TS_{\text{local}} = -\Delta\chi^{2} = -(\chi^{2}_{1}-\chi^{2}_{0})$, where $\chi^{2}_{1}$ and $\chi^{2}_{0}$ are respectively the $\chi^2$ values obtained when fitting the data with the alternative hypothesis and with the null hypothesis. The value of $2.71$ to set the upper limit at $95\%$ CL on the intensity of the feature, i.e. $s$, comes from the assumption that the two models differ by one free parameter. Under this assumption, the $TS_{\text{local}}$ should be distributed as a $\chi^{2}$ with one degree of freedom random variable (Wilks' theorem~\cite{wilks1938}). Indeed we see that, for both the box and line features, the $TS_{\text{local}}$ obeys this model in the high energy windows, while some deviations are observed at low energies. We have therefore implemented an alternative approach to evaluate the upper limits on the feature intensity, which takes into account the actual $TS_{\text{local}}$ distributions. We see that the results obtained with this approach do not differ significantly from those obtained with the ``standard'' approach based on Wilks' theorem, and therefore we have decided to quote the limits evaluated with the ``standard'' procedure.

Finally we evaluate the expectation bands for our results, i.e. the sensitivity to the null hypothesis, using a pseudo-experiment technique. We start fitting the observed CRE count distributions from the Sun and the anti-Sun with a simple PL model in the whole energy range, and we use this model as a template to evaluate the expected counts in each energy bin. Starting from the template model, a set of 1000 pseudo-experiments is performed, in which the counts in each energy bin are extracted from a Poisson distribution with mean value $\mu$ taken from the template, after adding a Gaussian fluctuation with $\sigma=f_{\text{syst}}~\mu$, where $f_{\text{syst}}$ is the fractional systematic uncertainty in the bin, to account for energy-dependent systematic uncertainties. The count distributions corresponding to the various pseudo-experiments are then fitted including the feature, and the containment bands (quantiles) for all the parameters are calculated. 

\section{Results}
\label{sec:res}

As discussed above, CREs originating from DM captured in the Sun should yield a signal peaked towards the direction of the Sun with an angular extension determined by the heliospheric magnetic field. A detailed study of the effects of the heliomagnetic field is beyond the goal of the present paper, and in general is not straightforward because the geometry of the field is rather complex. However, we have implemented a custom simulation of some specific field configurations based on the Parker model~\cite{Parker:1958zz} to calculate the trajectories of CREs produced between the Sun and the Earth. The simulation shows that the angular separation of the CRE directions at Earth from the Sun's direction depends on the CRE energy, on the production point of the particle and on the position of the Earth along its orbit; in particular, we find that for CREs with energies in the range considered for our analysis the maximum angular separations from the Sun are of a few tens degrees and decrease with increasing energy. Therefore, to take into account the effect of heliomagnetic field, in the present analysis we will use different RoIs with apertures of $10\degrees$, $30\degrees$ and $45\degrees$. However, we have also used smaller RoIs of $2\degrees$ and $5\degrees$ since CREs with energies above a few hundreds $\units{GeV}$ travelling from the Sun to the Earth are deflected of a few degrees~\cite{Bartoli:2019xvu}. The choice of the largest RoI will result in conservative limits.

\begin{figure}[!ht]
\centering
\includegraphics[width=\columnwidth,height=0.25\textheight,clip]{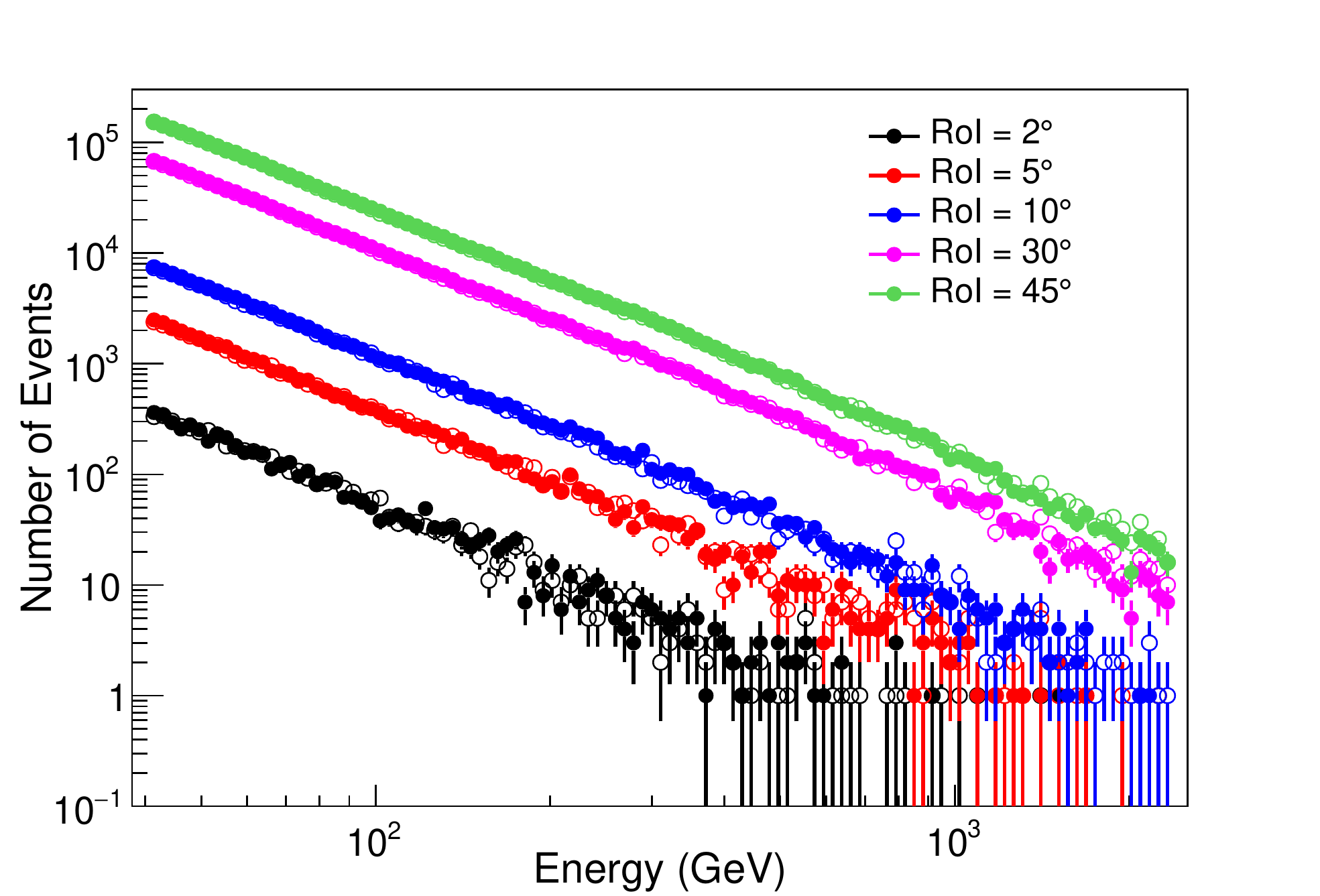}
\caption{Number of events as a function of the observed energy for RoIs of $2\degrees$ (black), $5\degrees$ (red), $10\degrees$ (blue), $30\degrees$ (magenta) and $45\degrees$ (green). Full circles: Sun RoIs; open circles: anti-Sun RoIs. The histograms are divided in 64 bins per energy decade. The error bars correspond to the square root of the number of events in each bin.}
\label{fig:counts}
\end{figure}

Figure~\ref{fig:counts} shows the number of events as a function of observed energy for the above RoIs towards the Sun and the anti-Sun. All distributions look featureless, and follow a power-law behavior.

\begin{figure*}[!hb]
\centering
\begin{tabular}{ccc}
\includegraphics[width=0.32\textwidth,height=0.18\textheight,clip]{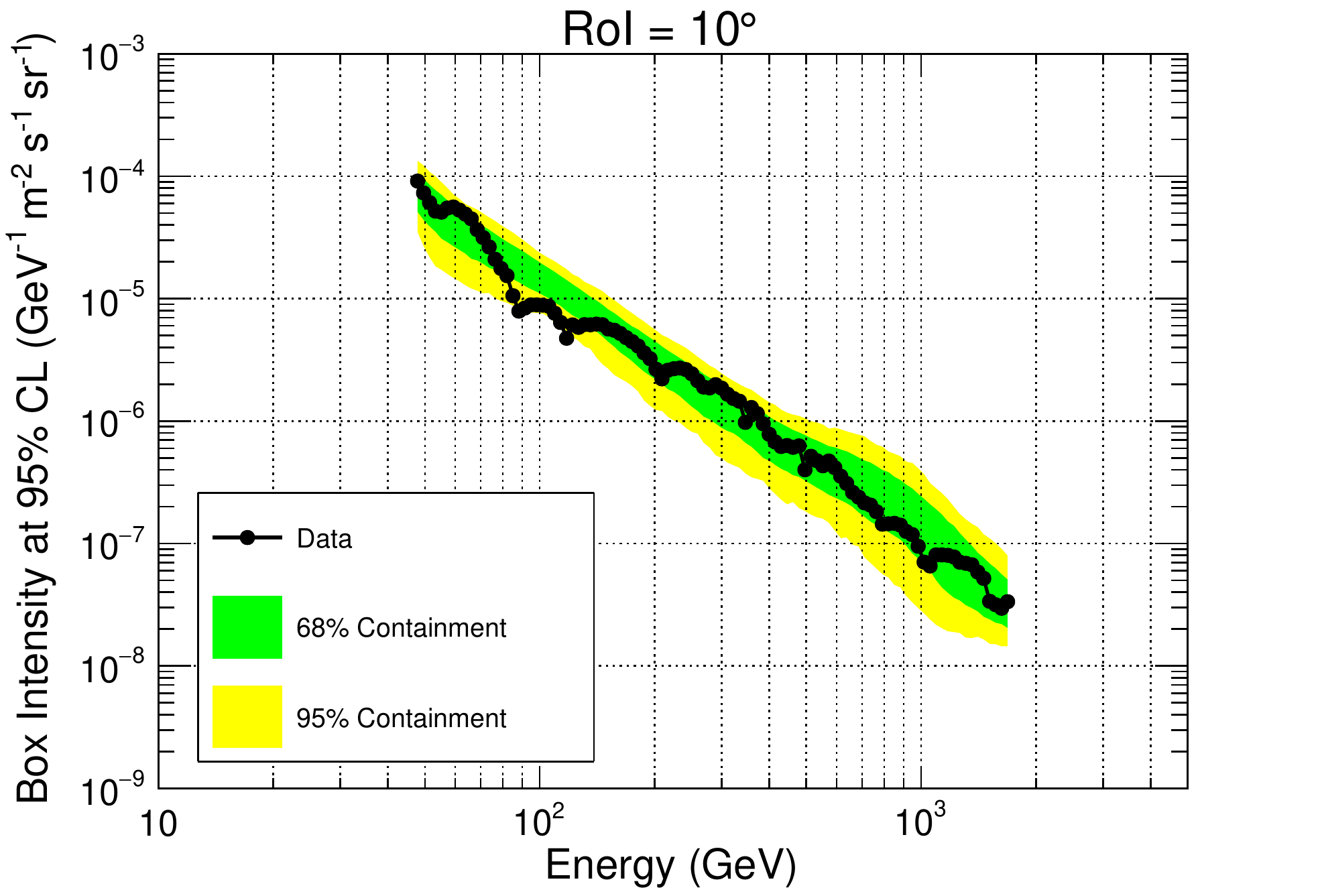} &
\includegraphics[width=0.32\textwidth,height=0.18\textheight,clip]{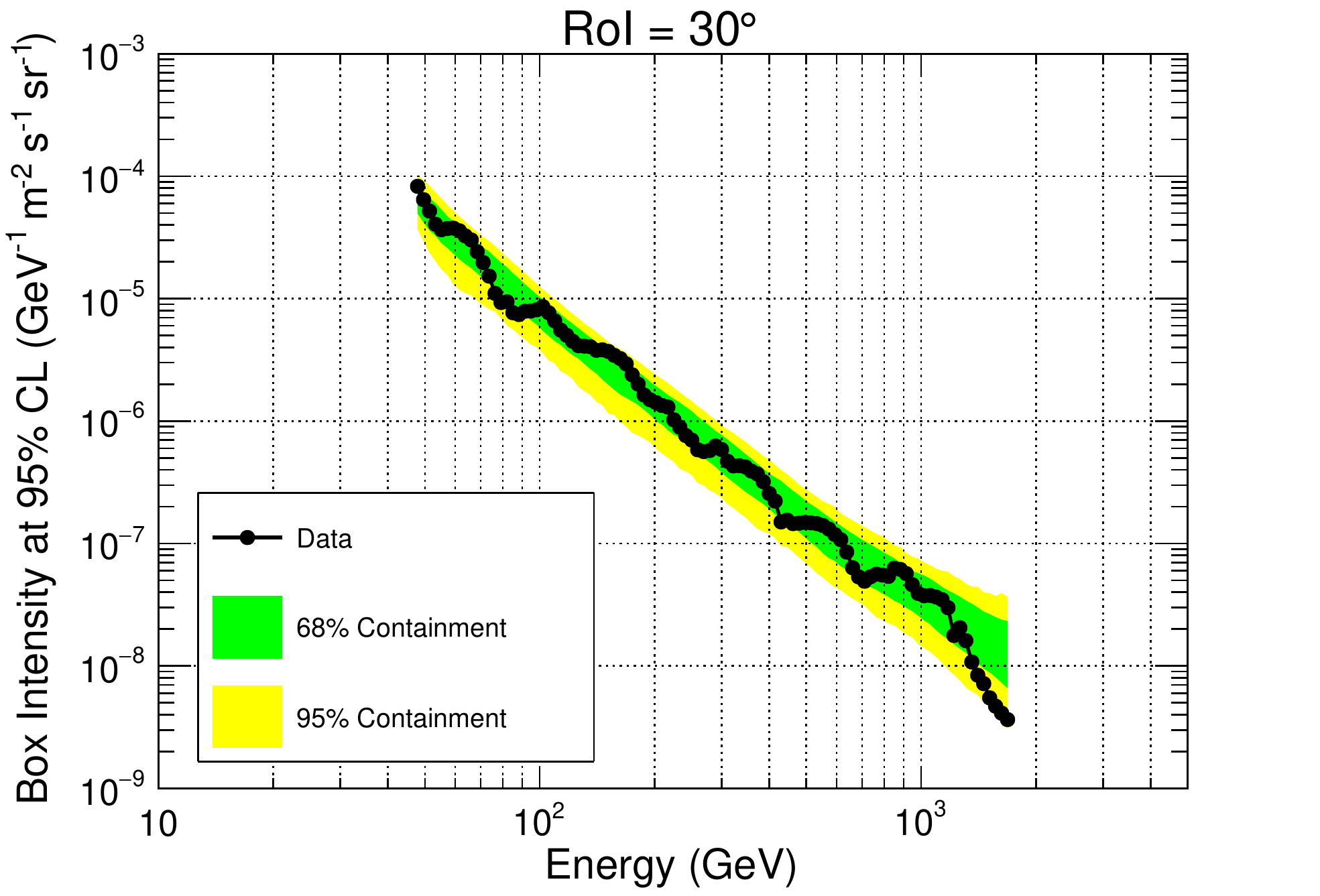} &
\includegraphics[width=0.32\textwidth,height=0.18\textheight,clip]{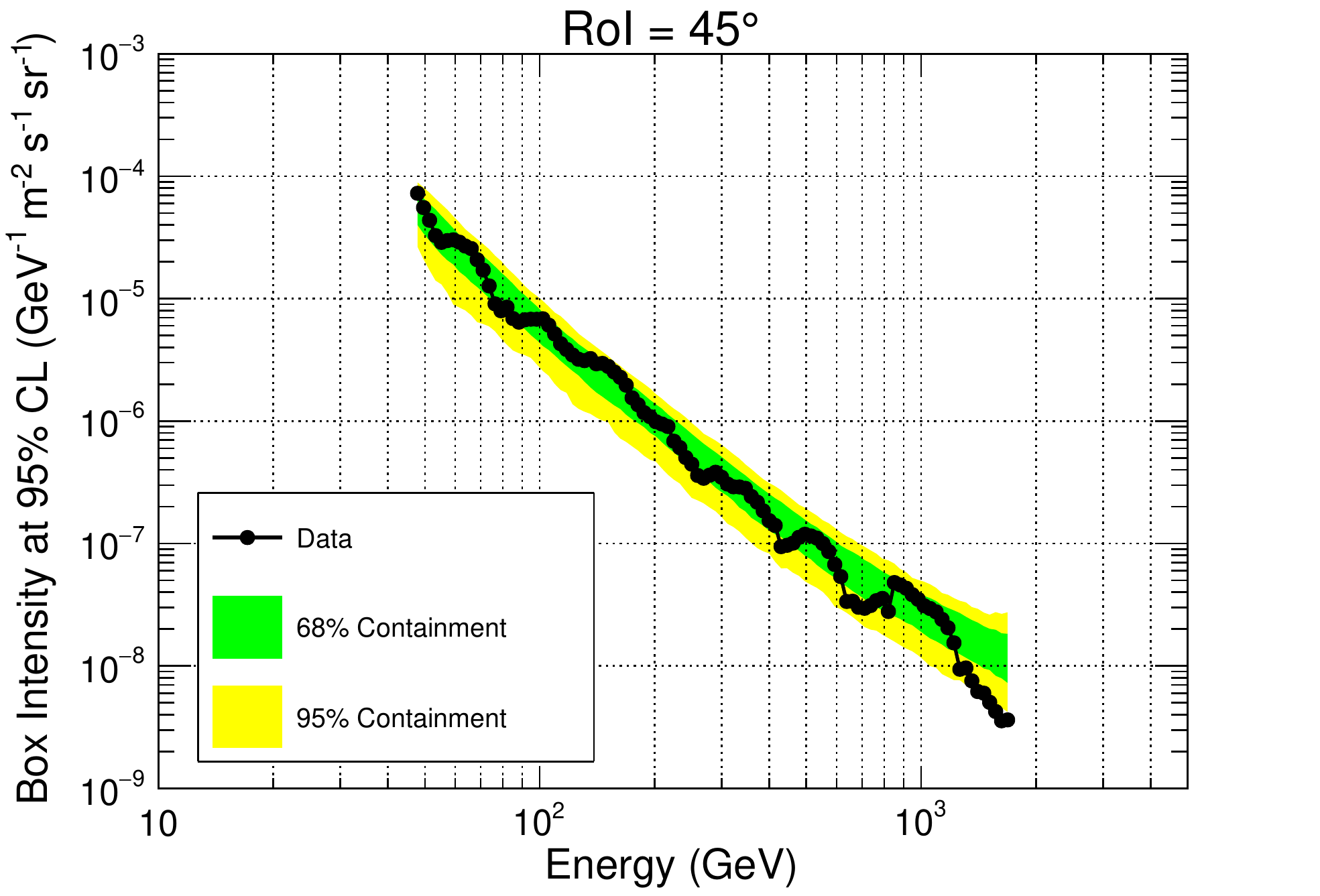} \\
\includegraphics[width=0.32\textwidth,height=0.18\textheight,clip]{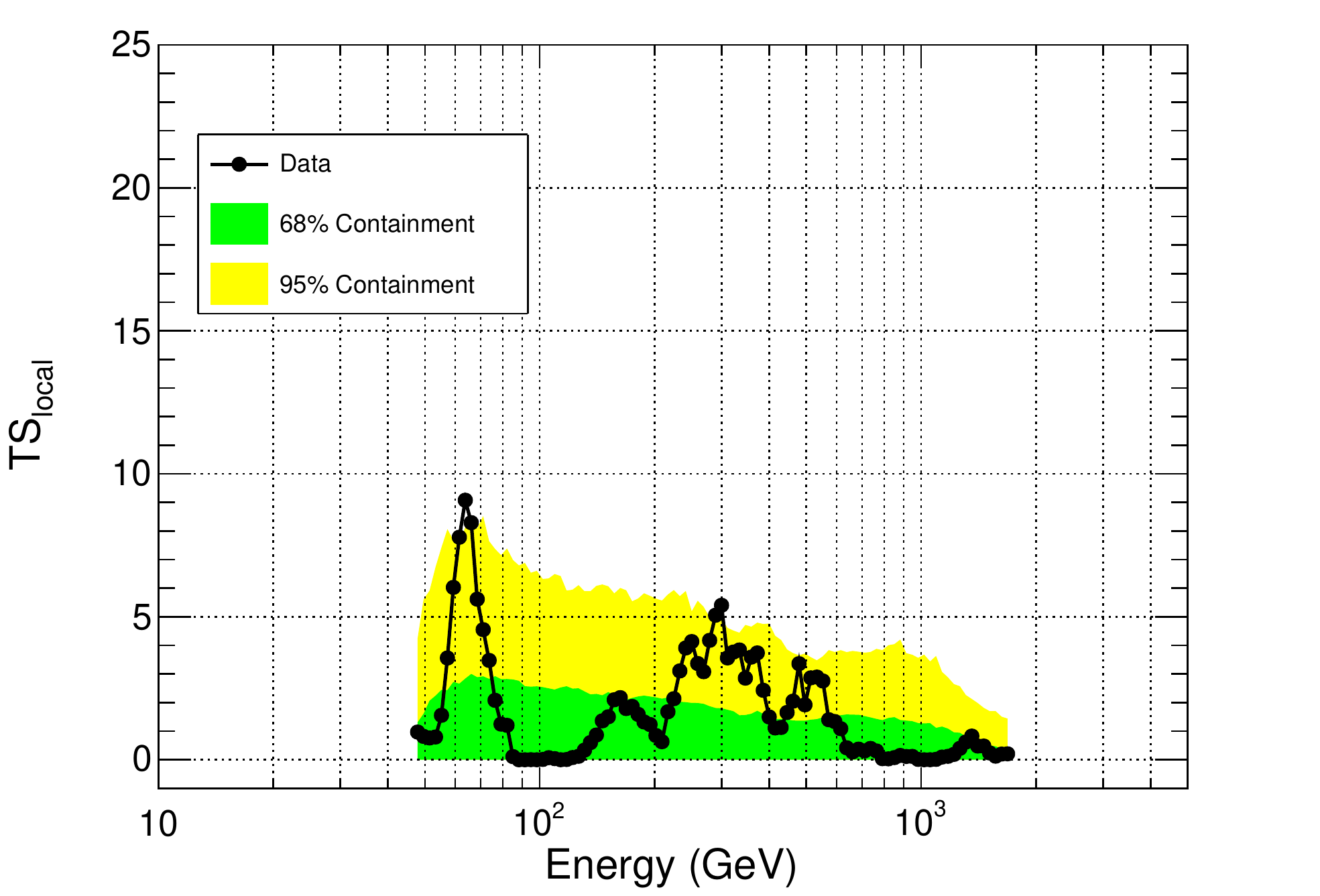} &
\includegraphics[width=0.32\textwidth,height=0.18\textheight,clip]{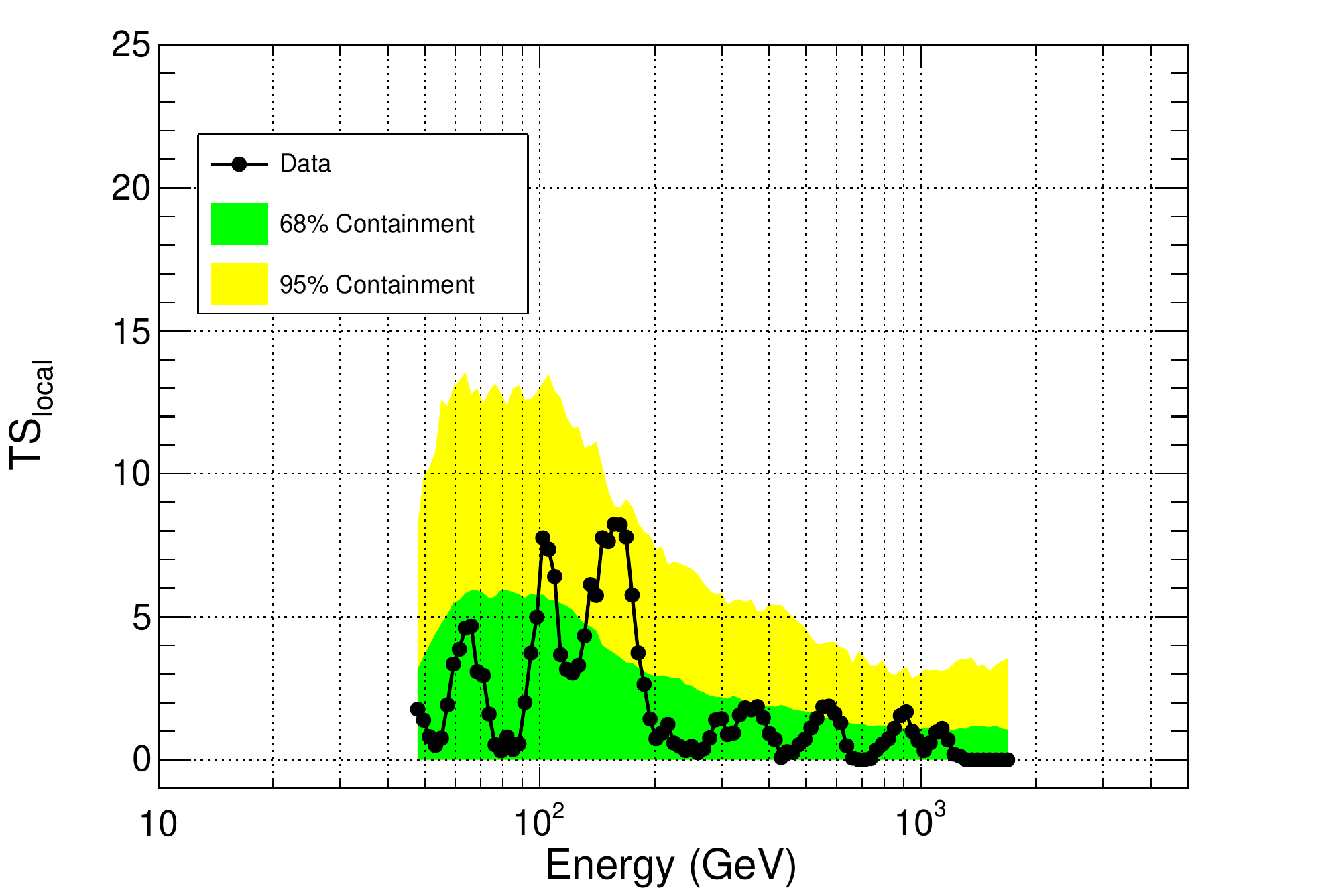} &
\includegraphics[width=0.32\textwidth,height=0.18\textheight,clip]{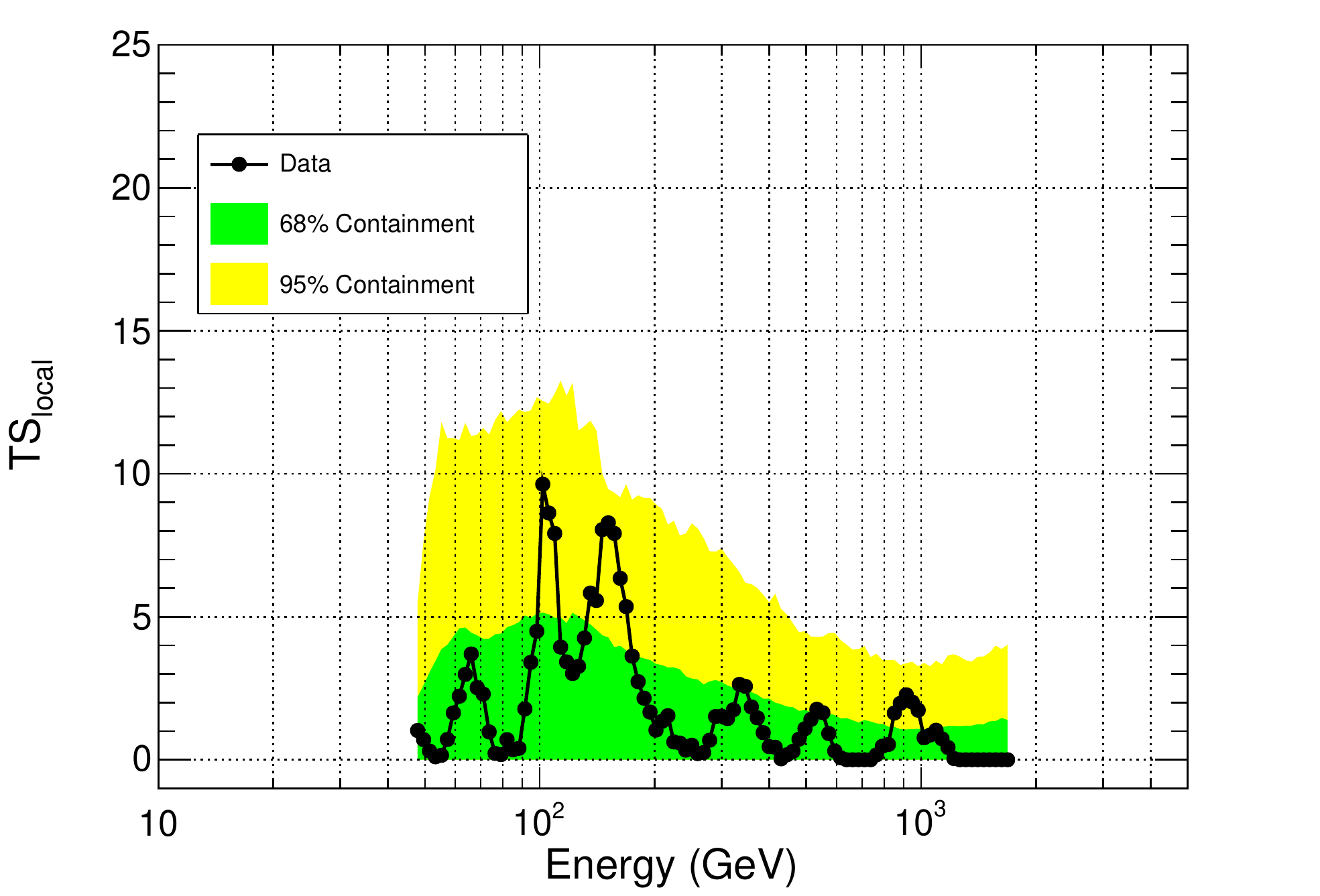} \\
\includegraphics[width=0.32\textwidth,height=0.18\textheight,clip]{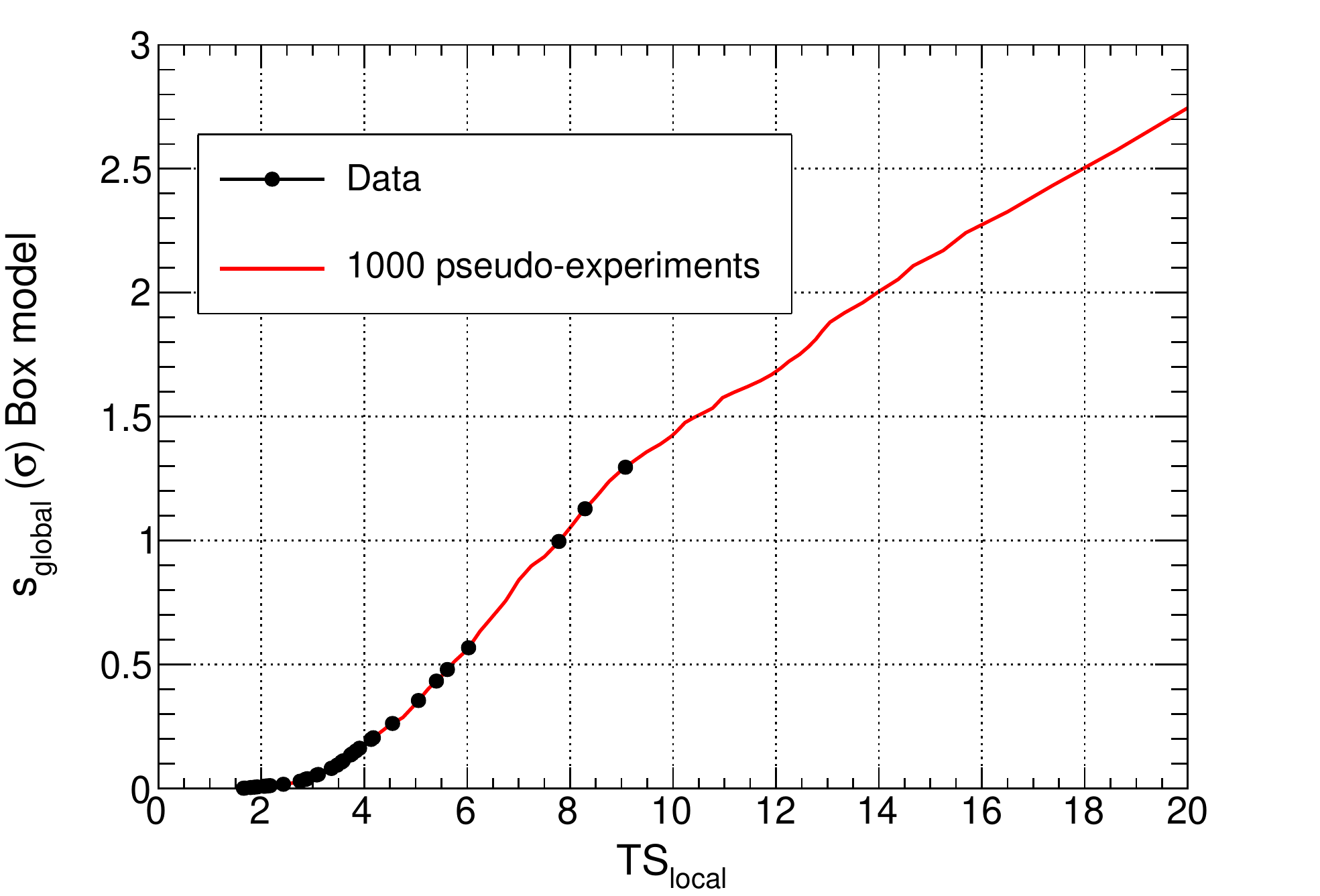} &
\includegraphics[width=0.32\textwidth,height=0.18\textheight,clip]{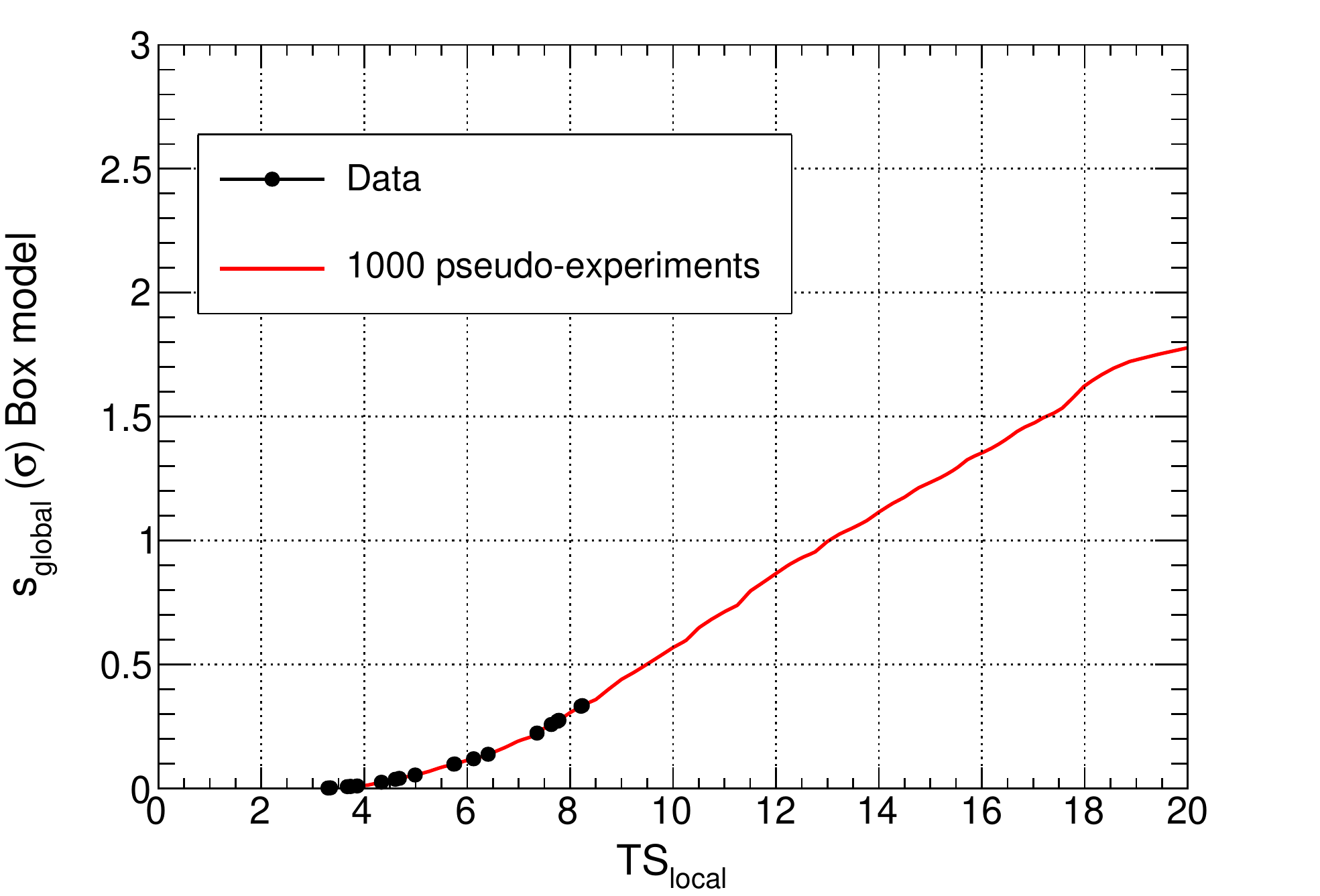} &
\includegraphics[width=0.32\textwidth,height=0.18\textheight,clip]{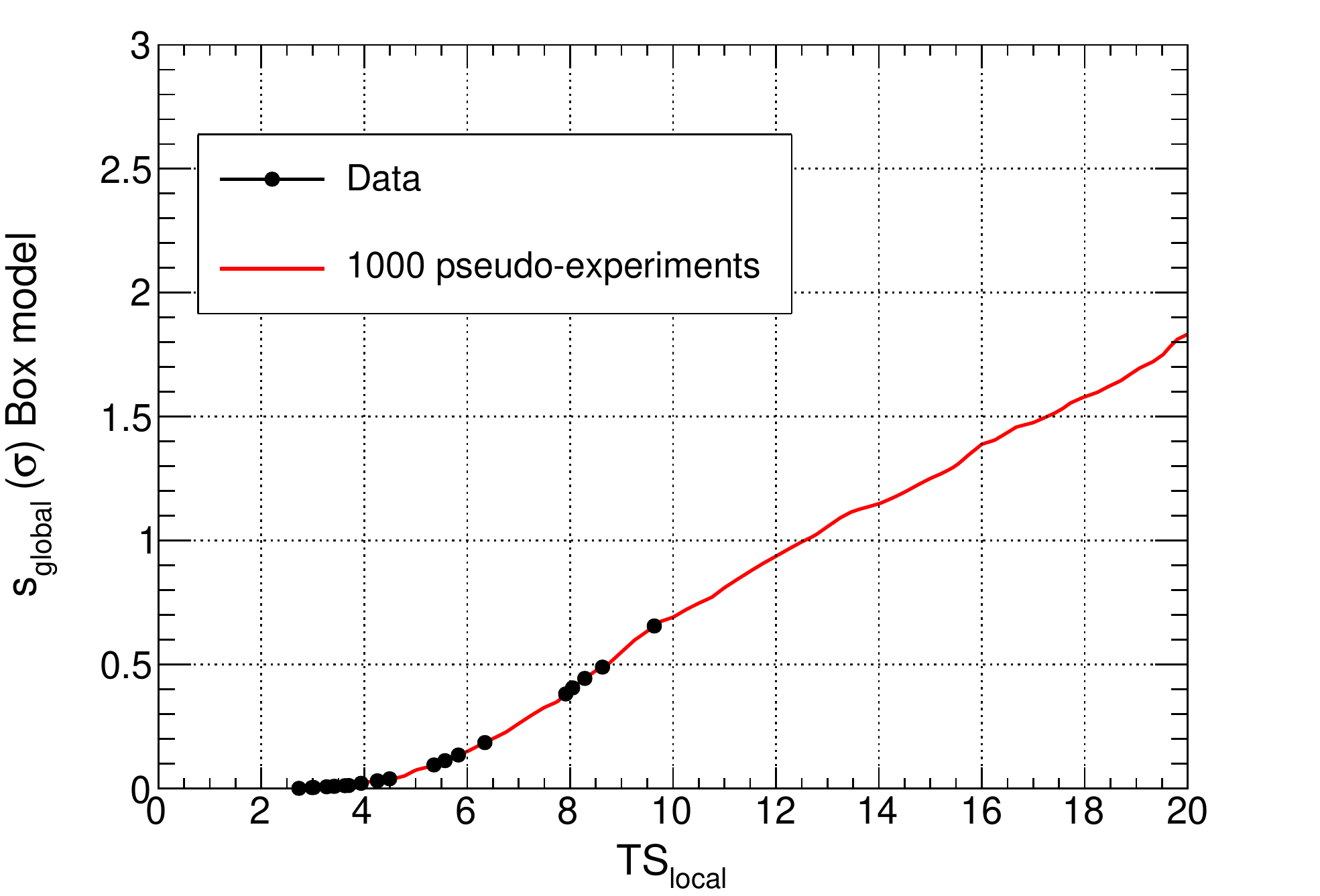} \\
\includegraphics[width=0.32\textwidth,height=0.18\textheight,clip]{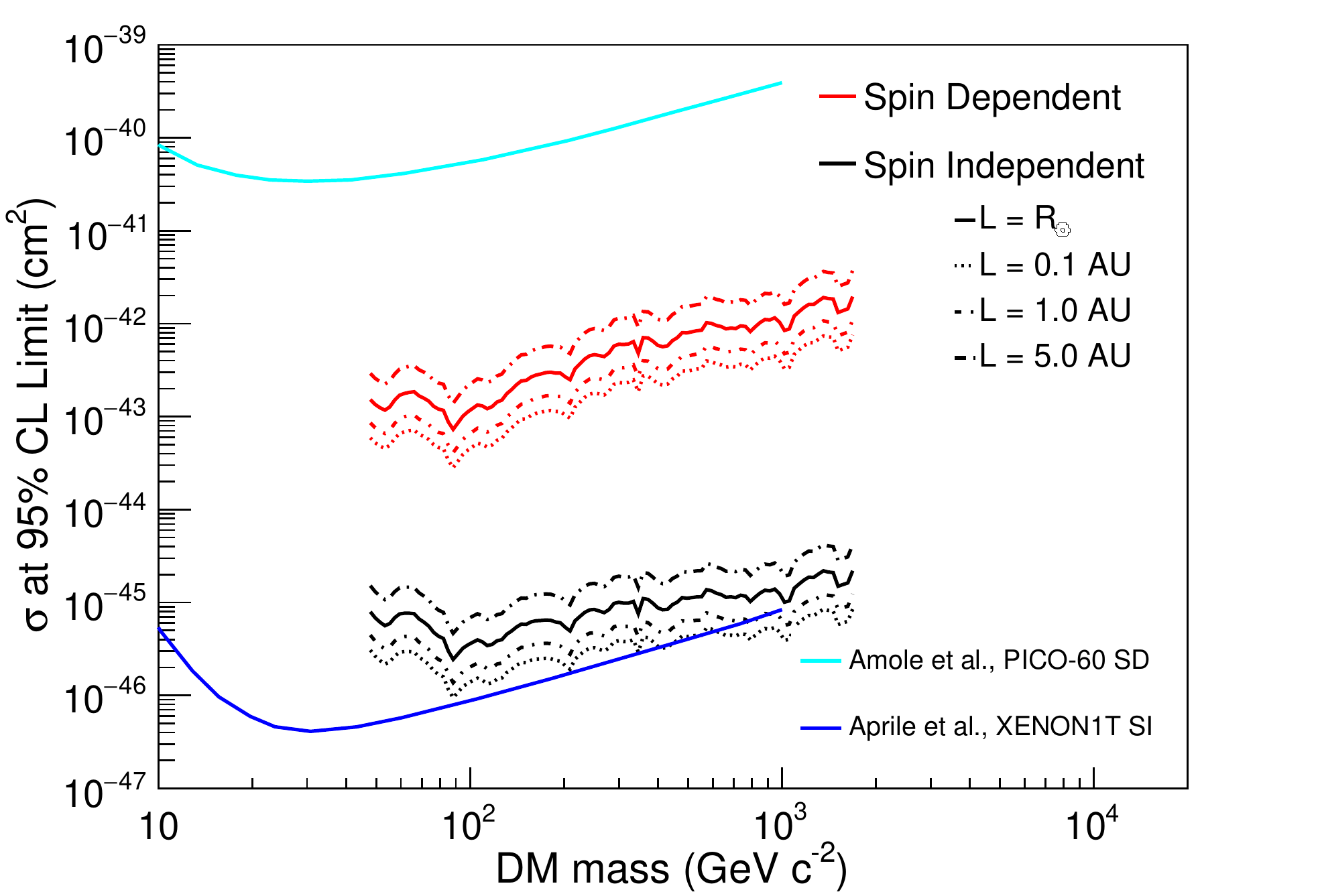} &
\includegraphics[width=0.32\textwidth,height=0.18\textheight,clip]{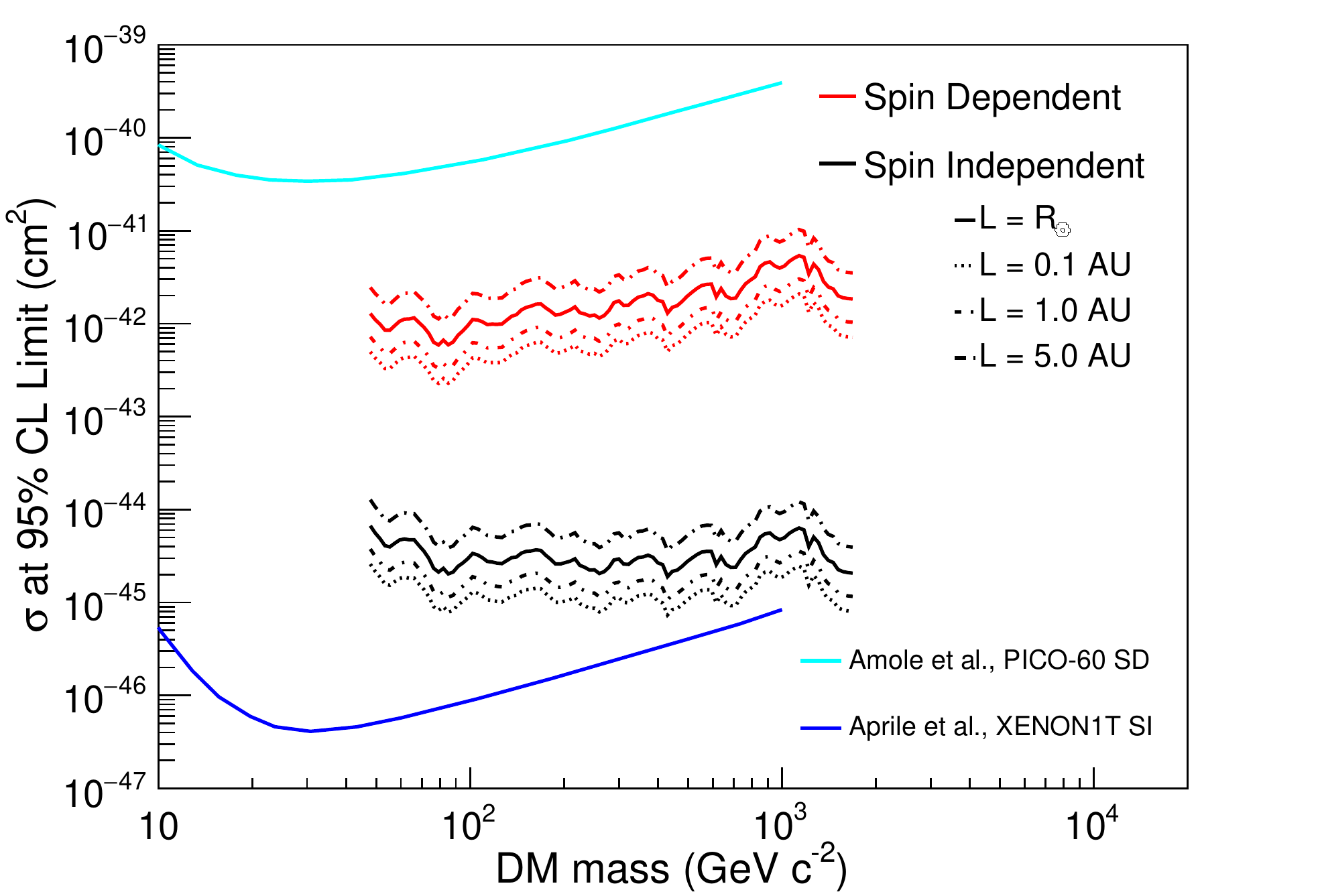} &
\includegraphics[width=0.32\textwidth,height=0.18\textheight,clip]{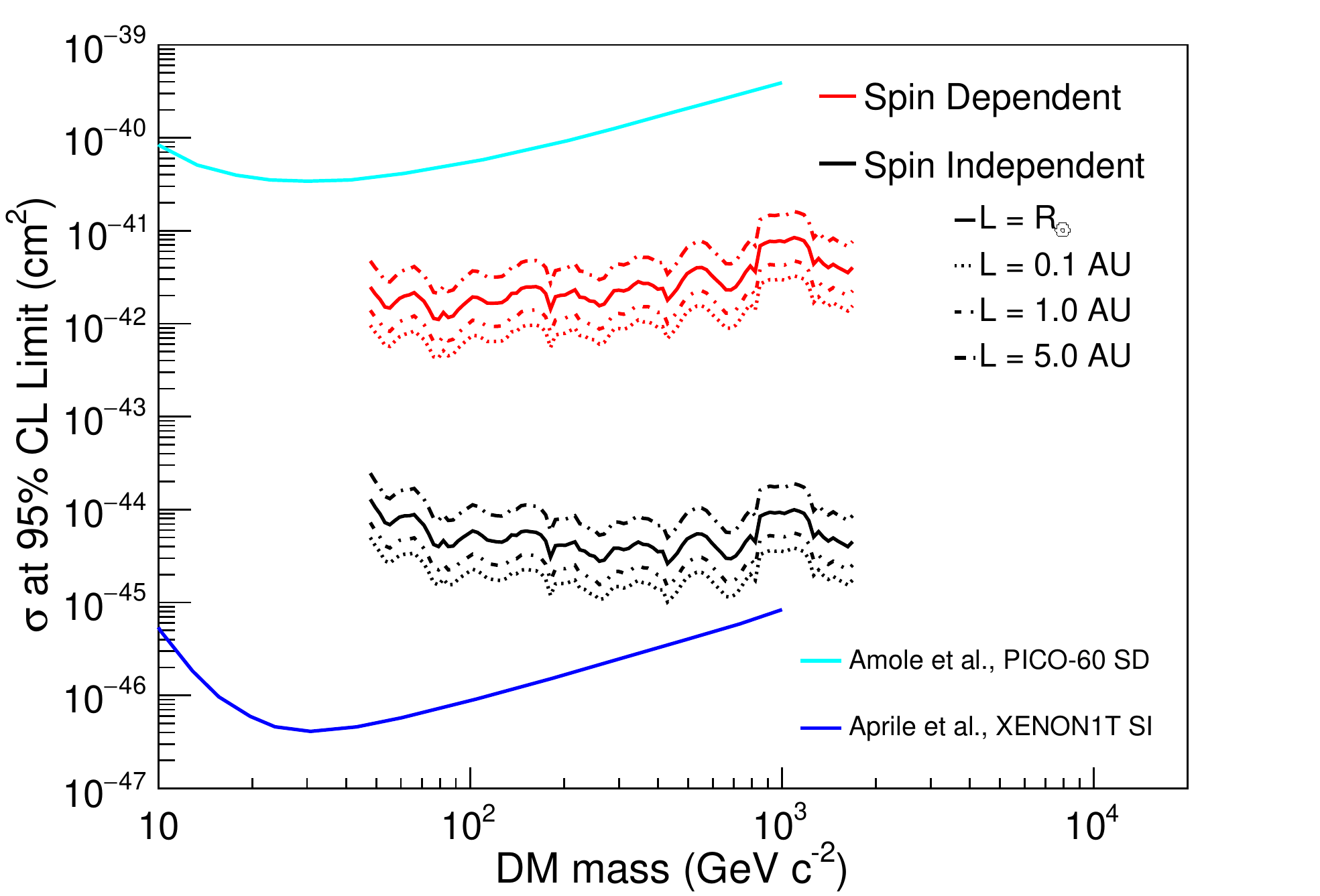}\\
\end{tabular}
\caption{Results of the search for box-like features in $10\degrees$ (left column), $30\degrees$ (middle column) and $45\degrees$ (right column) RoIs. The first row shows the upper limit at $95\%$ CL on the line intensity compared with the expectations from the pseudo-experiments (the green and yellow bands show the central $68\%$ and $95\%$ expectation bands for the $95\%$ CL limit). The second row shows the value of the local TS in the various energy windows compared with the expectations from the pseudo-experiments (the green and yellow bands show the one-sided $68\%$ and $95\%$ expectation bands for the local TS). The third row shows the conversion from the local TS to the global significance. The last row shows the upper limits on the DM-nucleon cross section for both the spin-dependent and spin-independent cases for four different decay length for the long-lived mediator $L=R_{\odot}$, $0.1$, $1$ and $5\units{AU}$. The plots also show the limits at $90\%$ CL from the PICO-60 experiment~\cite{Amole:2019fdf} in the case of spin-dependent (SD) scattering (cyan line) and from the XENON1T experiment~\cite{Aprile:2018dbl} in the case of spin-independent (SI) scattering (blue line).}
\label{fig:box}
\end{figure*}

\medskip

Figure~\ref{fig:box} shows the results obtained when using the box model for the features (i.e. the light intermediate state model). The results for the $10\degrees$, $30\degrees$ and $45\degrees$ RoIs are shown in the left, middle and right column, respectively. The upper limits (ULs) on the intensity of the box-like feature at 95\% confidence level (CL) are shown in the first row (starting from the top). The plots also show the central 68\% and 95\% containment belts for the expected limits evaluated with the pseudo-experiment technique described in Sec.~\ref{sec:anamet}. In most cases the measured limits lie within the 95\% containment belt.

The second row of Figure~\ref{fig:box} shows the local significance of possible features at different energies. The plots also show the one-sided $68\%$ and $95\%$ expectation bands obtained with the pseudo-experiment technique for the local TS. In most energy windows, the values of $TS_{\text{local}}$ lie within the $95\%$ expectation band. Nonetheless, there are some fits yielding values of $TS_{\text{local}}$ slightly above the $95\%$ expectation bands. However, in the evaluation of the global significance of these possible features, it should be kept in mind that the fits are not independent and the number of trials should be taken into account. As a consequence, possible features associated with a local significance up to about $2 \sigma$ turn out to be globally insignificant. To calculate the global significances we use the 1000 pseudo-experiments discussed in Sec.~\ref{sec:anamet}. For each pseudo-experiment (which corresponds to a simulation of one full search across the entire energy range) we record the largest value of the local Test Statistic, $TS_{\rm max}$. We then calculate the quantiles of the distribution of $TS_{\rm max}$ and we convert our global significance $s_{\text{global}}$ to a corresponding number of sigmas assuming that $s_{\text{global}}$ obeys a (half) normal distribution. The third row of Figure~\ref{fig:box} shows the conversion from $TS_{\text{ local}}$ to $s_{\text{global}}$ models. The most significant features have global significances less than 1.5$\sigma$.

Finally, starting from the ULs of the first row and using eq.~\ref{eq:phidm} with the assumption that the capture rate scales linearly with the DM-nucleon cross section, we evaluate the limits of the DM-nucleon cross-section, as:

\begin{equation}
    \sigma_{\text{UL}}(m_\chi) = \frac{I_{\text{UL}}(E=m_\chi)~\Delta\Omega}{\Phi_{\text{DM}}(E=m_\chi)} \times 10^{-40}~\units{cm^2}
\end{equation}
where $\Delta\Omega$ is the solid angle corresponding to the selected RoI, $\Phi_{\text{DM}}(E)$ is taken from Eq.~\ref{eq:phidm}, and the value of $10^{-40} \units{cm^2}$ accounts for the DM-nucleon cross section value used to calculate the capture rate shown in Fig.~\ref{fig:fig2}. 

The fourth row in Figure~\ref{fig:box} shows the constraints on DM annihilation to $e^+e^-$ via an intermediate state at 95\% CL, obtained from the upper limits on the solar CRE intensity, assuming that the capture of DM takes place either via spin-independent scattering (black lines) or spin-dependent scattering (red lines). The constraints have been calculated for four values of the decay length of the intermediate state, $L=R_{\odot}$, $0.1$, $1$ and $5\units{AU}$. The limits on $\sigma_{SI}$ are in the range from about $5 \times 10^{-47} \units{cm^{2}}$ to about $5 \times 10^{-45} \units{cm^{2}}$, depending on the RoI and on the decay length of the mediator. For$\units{TeV}$ DM masses the limits are similar for all the RoIs, while for DM masses around $100 \units{GeV}$ they are stronger for the smallest RoI, as expected. The limits on $\sigma_{\text{SD}}$ exhibit the same behavior as those on $\sigma_{\text{SI}}$, although they are up to three orders of magnitude higher, as expected from the values of the capture rates (see Fig.~\ref{fig:fig2}). The limits at $90\%$ CL from the PICO-60 experiment~\cite{Amole:2019fdf} in the case of spin-dependent scattering and from the XENON1T experiment ~\cite{Aprile:2018dbl} in the case of spin-independent scattering are also shown. Our limits are consistent with those obtained from DM direct search experiments in the same mass range~\cite{Amole:2019fdf,Aprile:2018dbl,Agnes:2018fwg,Akerib:2017kat}.
However, we note that XENON1T and PICO-60 operate with target materials and under astrophysical time-scales different from what is used for the DM capture in the Sun.

The limits derived here are about two or more orders of magnitude weaker than those derived in Ref.~\cite{Ajello:2011dq}. The main difference with respect to~\cite{Ajello:2011dq} is that in the present work the expected CRE fluxes from DM annihilations are calculated with Eq.~\ref{eq:phidm} (point-like approximation),
while in~\cite{Ajello:2011dq} the extension of the signal was taken into account (see Eq.~21 in Ref.~\cite{Ajello:2011dq}, which in turn is based on Eq.~4 of Ref.~\cite{Schuster:2009fc}). 
The current model is the same employed by several authors in their recent works~\cite{Albert:2018jwh,Leane:2017vag} and should yield the same results starting from Eq.~4 in Ref.~\cite{Schuster:2009fc} in the relativistic limit, i.e. $m_{\phi} \ll m_{\chi}$. 
To check more in detail this aspect, we have  implemented a full Monte Carlo simulation of the whole 3D propagation and decaying process. We find, indeed, that the two methods give indistinguishable results, validating the accuracy of the point-like approximation.
In turn, this highlights that a problem was present in the analytic calculation of the flux via Eq.~21 in~\cite{Ajello:2011dq}. 

The decays of the mediator to $e^\pm$ can also produce photons in the final state (final state radiation, FSR). Recently, limits on $\sigma_{\text{SD}}$ were obtained from from $\units{TeV}$ observations of the Sun with the HAWC detector~\cite{Albert:2018jwh}. Our limits are competitive with those limits in the $\units{TeV}$ DM mass range.  

\begin{figure*}[!hb]
\centering
\begin{tabular}{ccc}
\includegraphics[width=0.32\textwidth,height=0.18\textheight,clip]{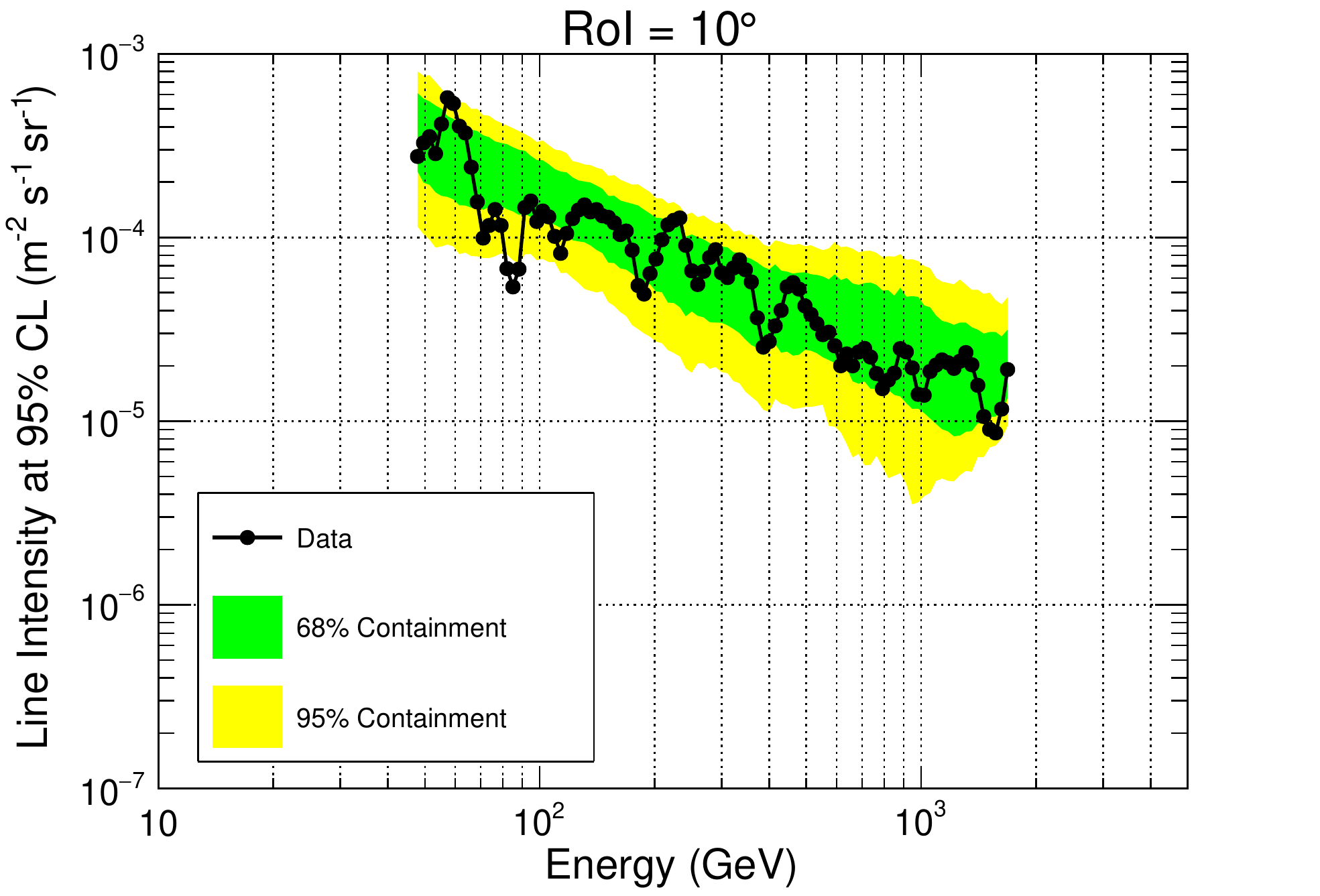} &
\includegraphics[width=0.32\textwidth,height=0.18\textheight,clip]{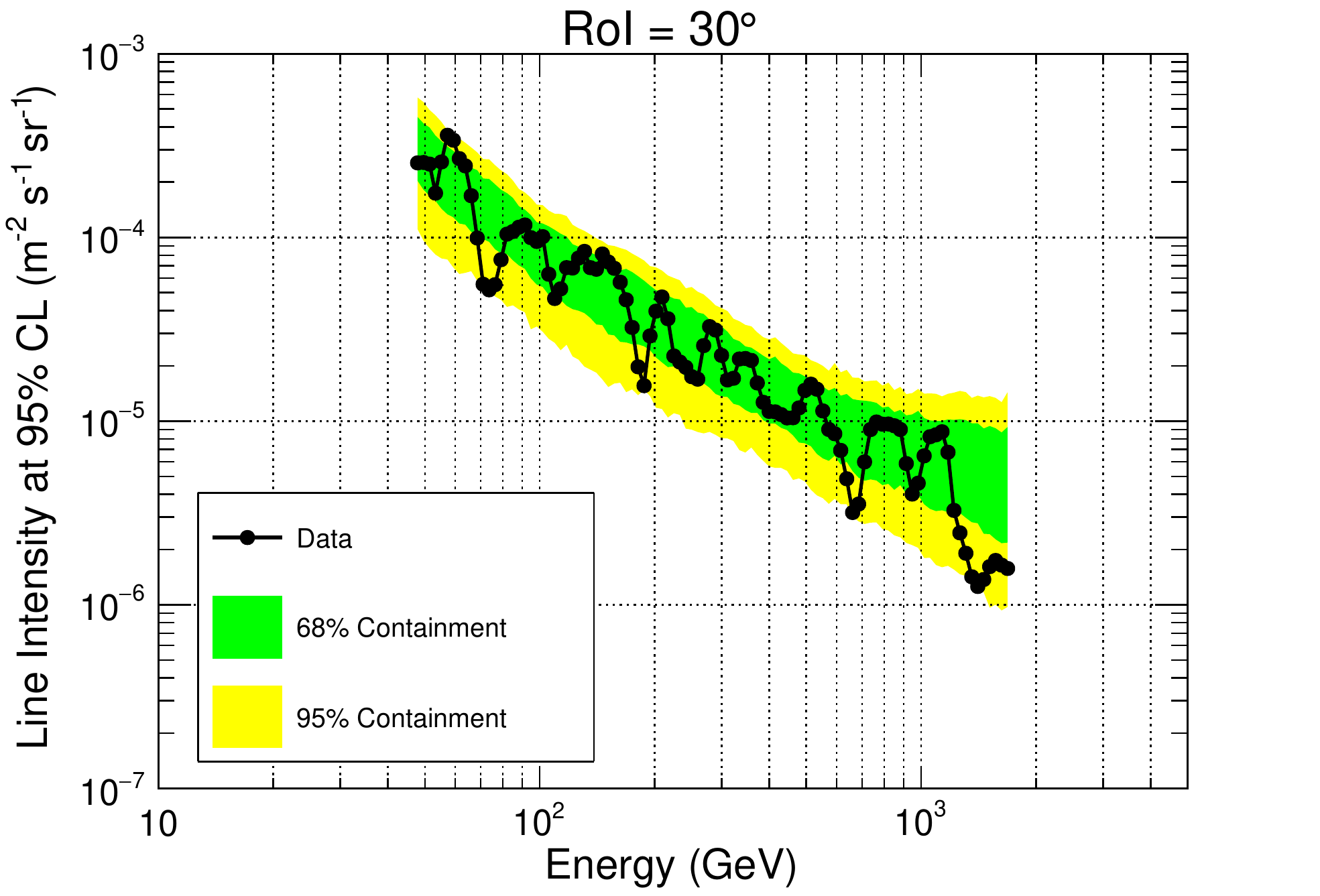} &
\includegraphics[width=0.32\textwidth,height=0.18\textheight,clip]{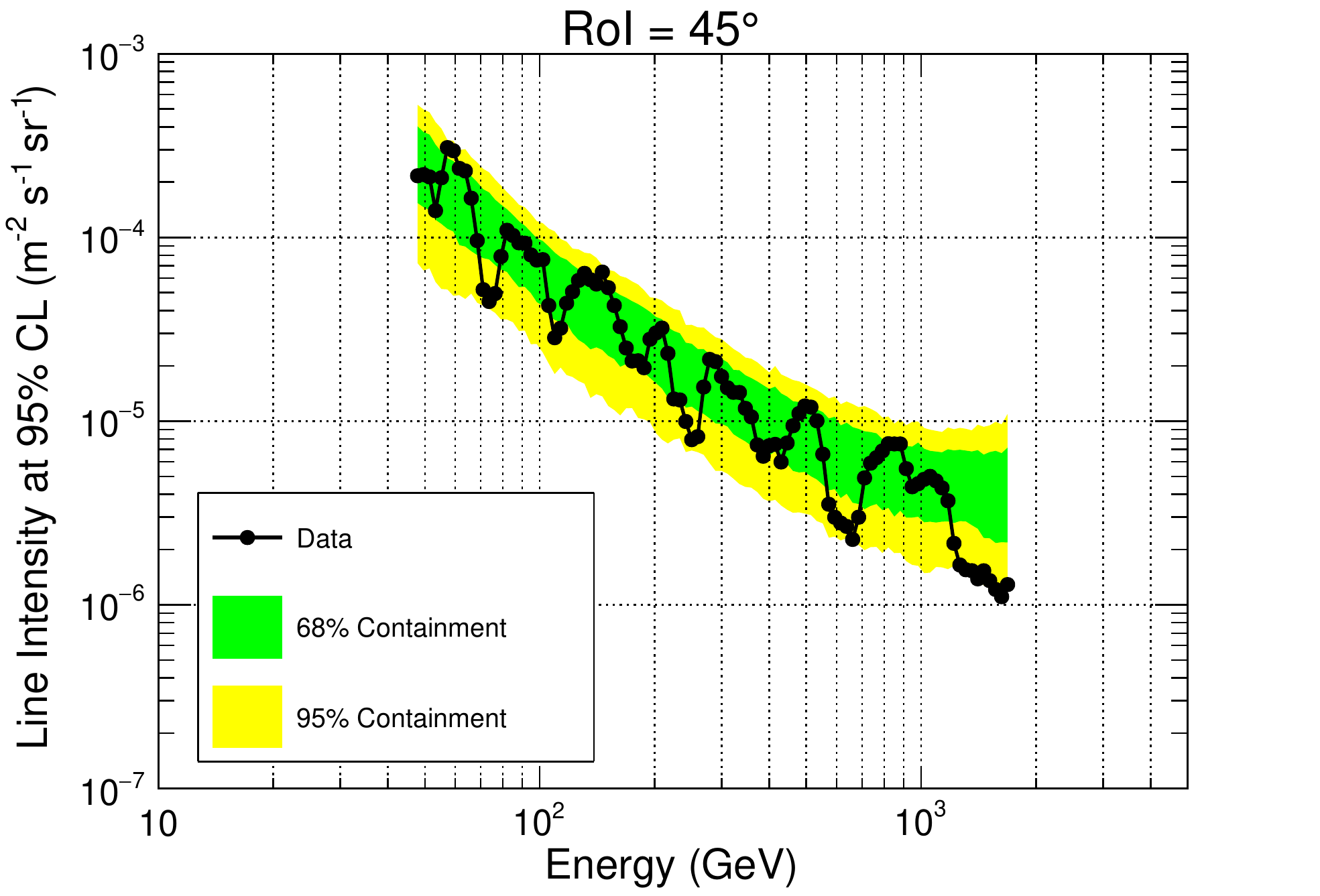} \\
\includegraphics[width=0.32\textwidth,height=0.18\textheight,clip]{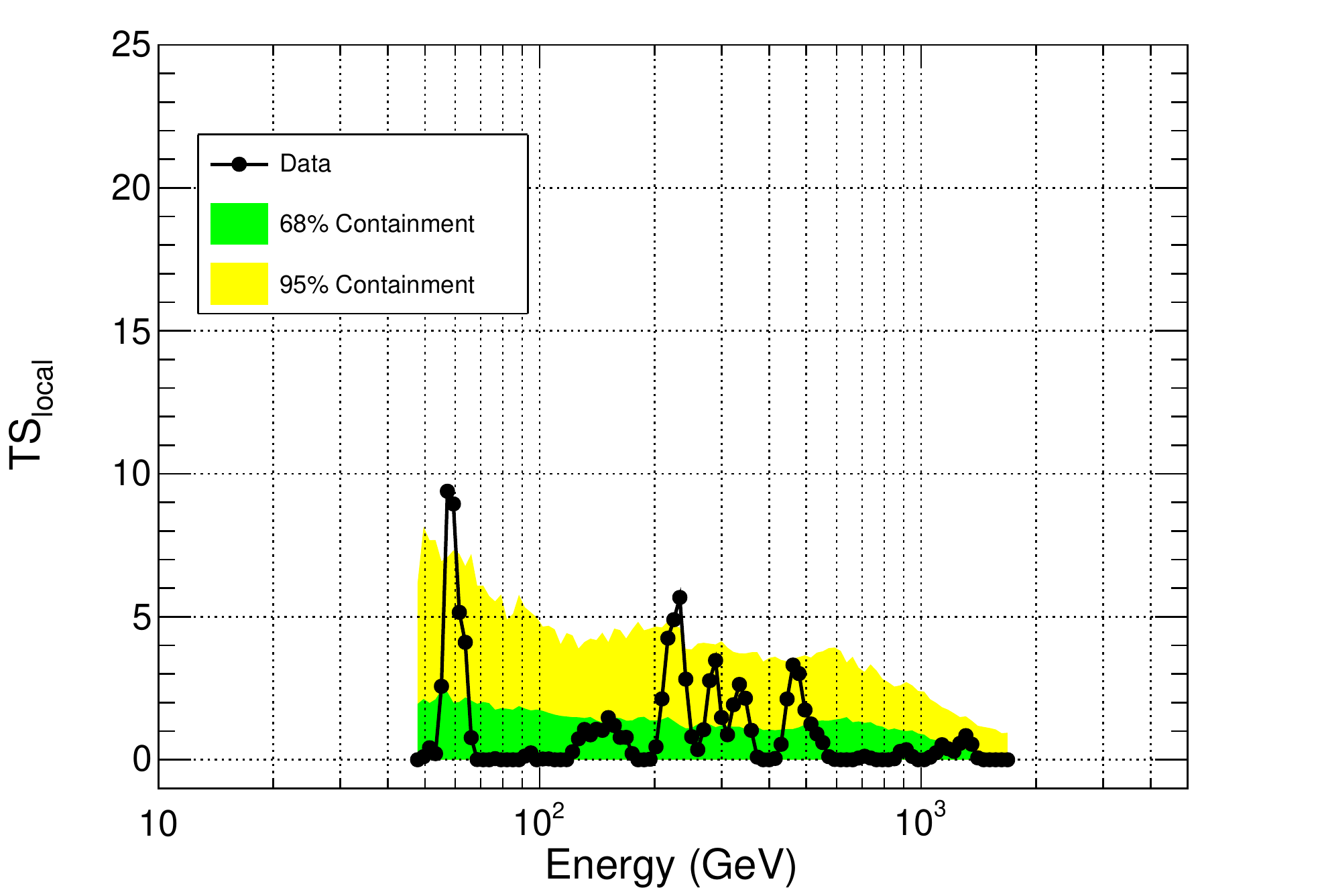} &
\includegraphics[width=0.32\textwidth,height=0.18\textheight,clip]{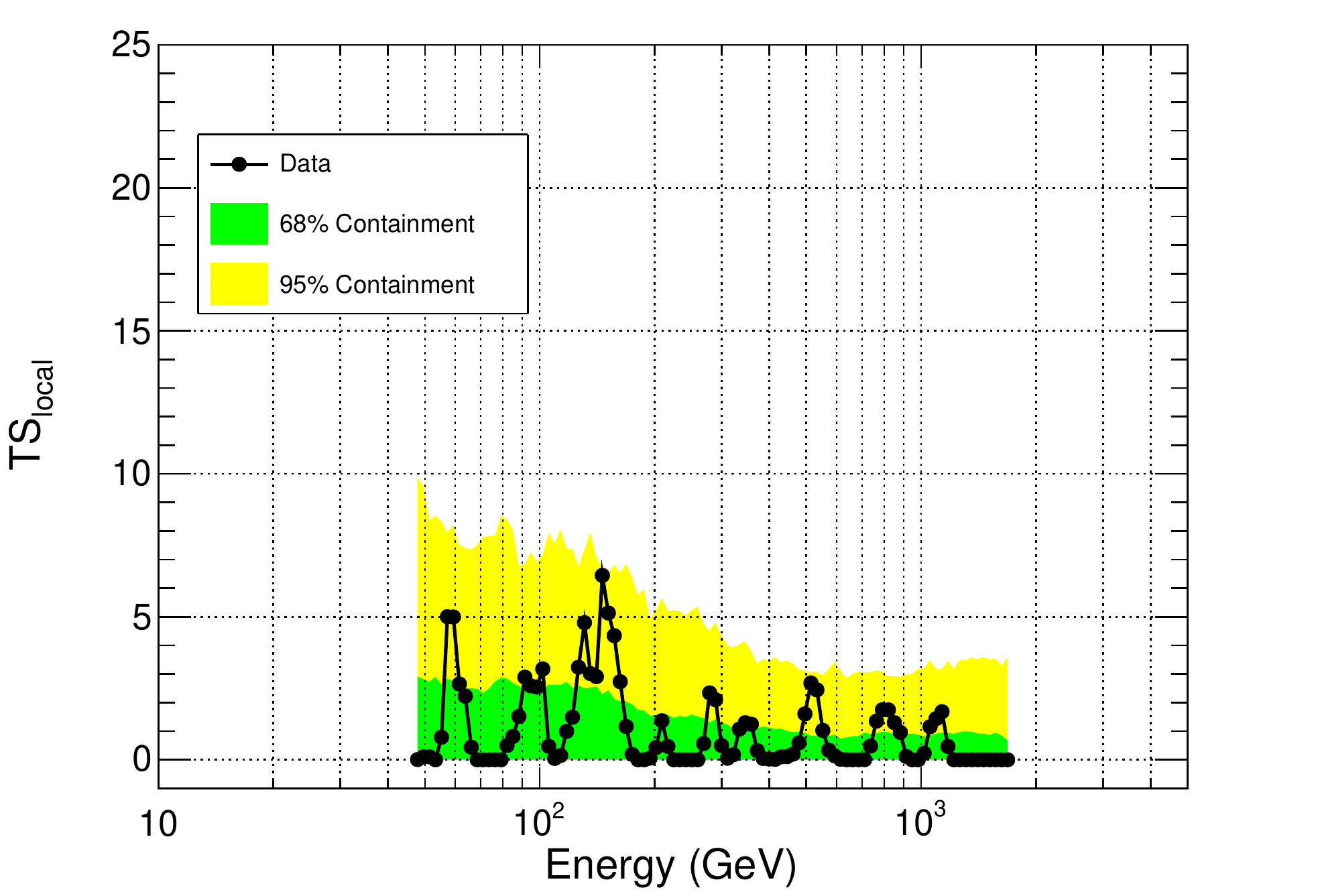} &
\includegraphics[width=0.32\textwidth,height=0.18\textheight,clip]{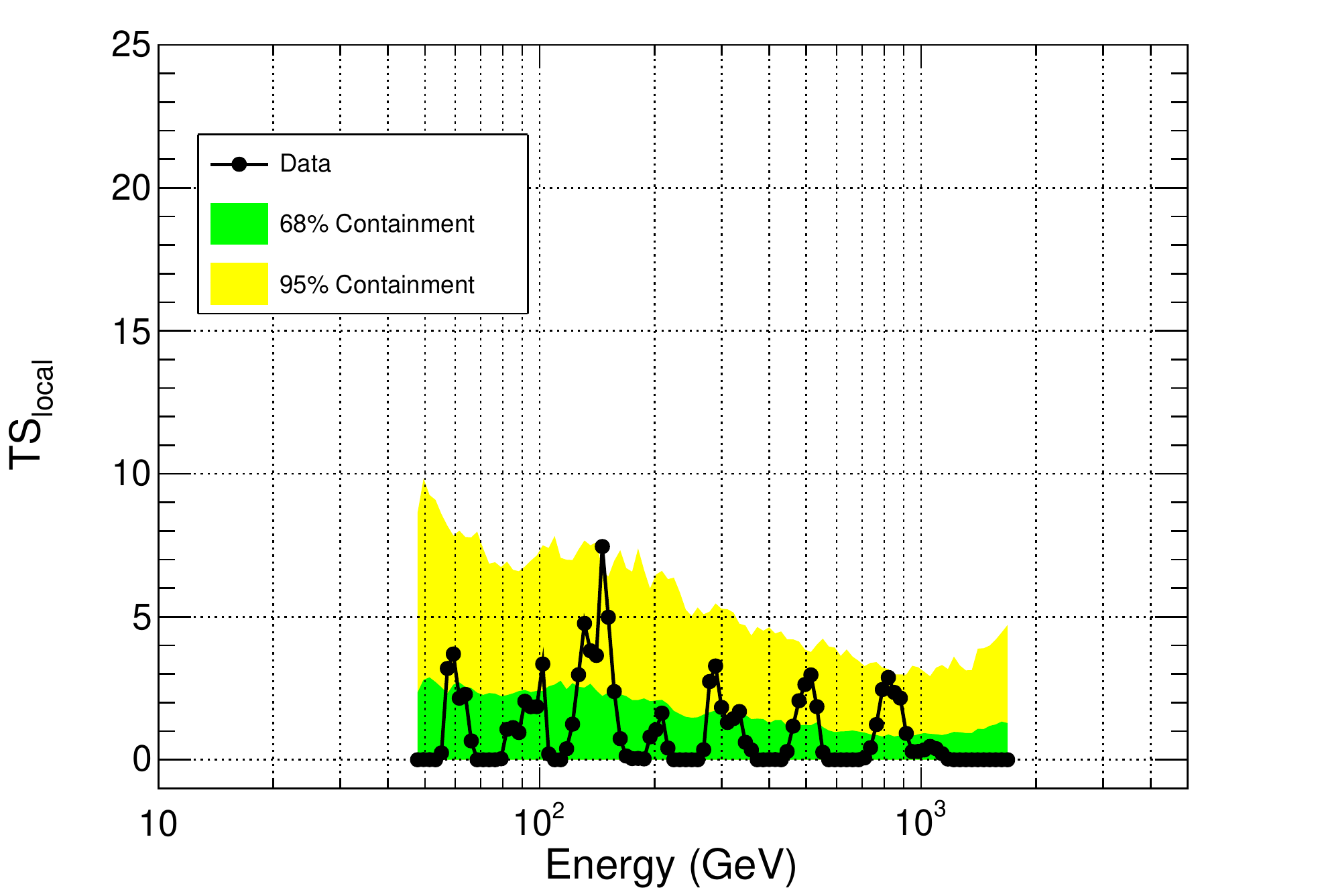} \\
\includegraphics[width=0.32\textwidth,height=0.18\textheight,clip]{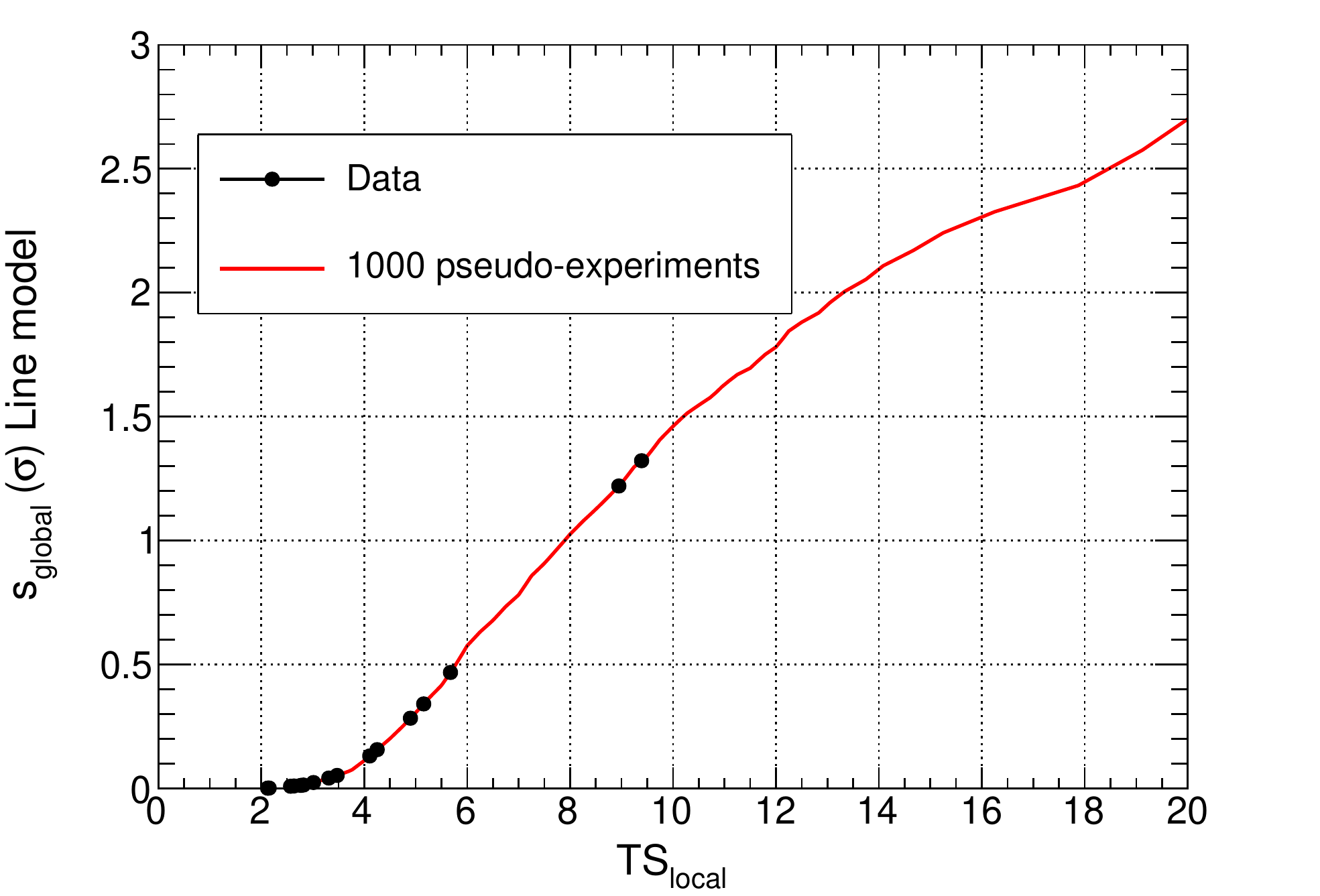} &
\includegraphics[width=0.32\textwidth,height=0.18\textheight,clip]{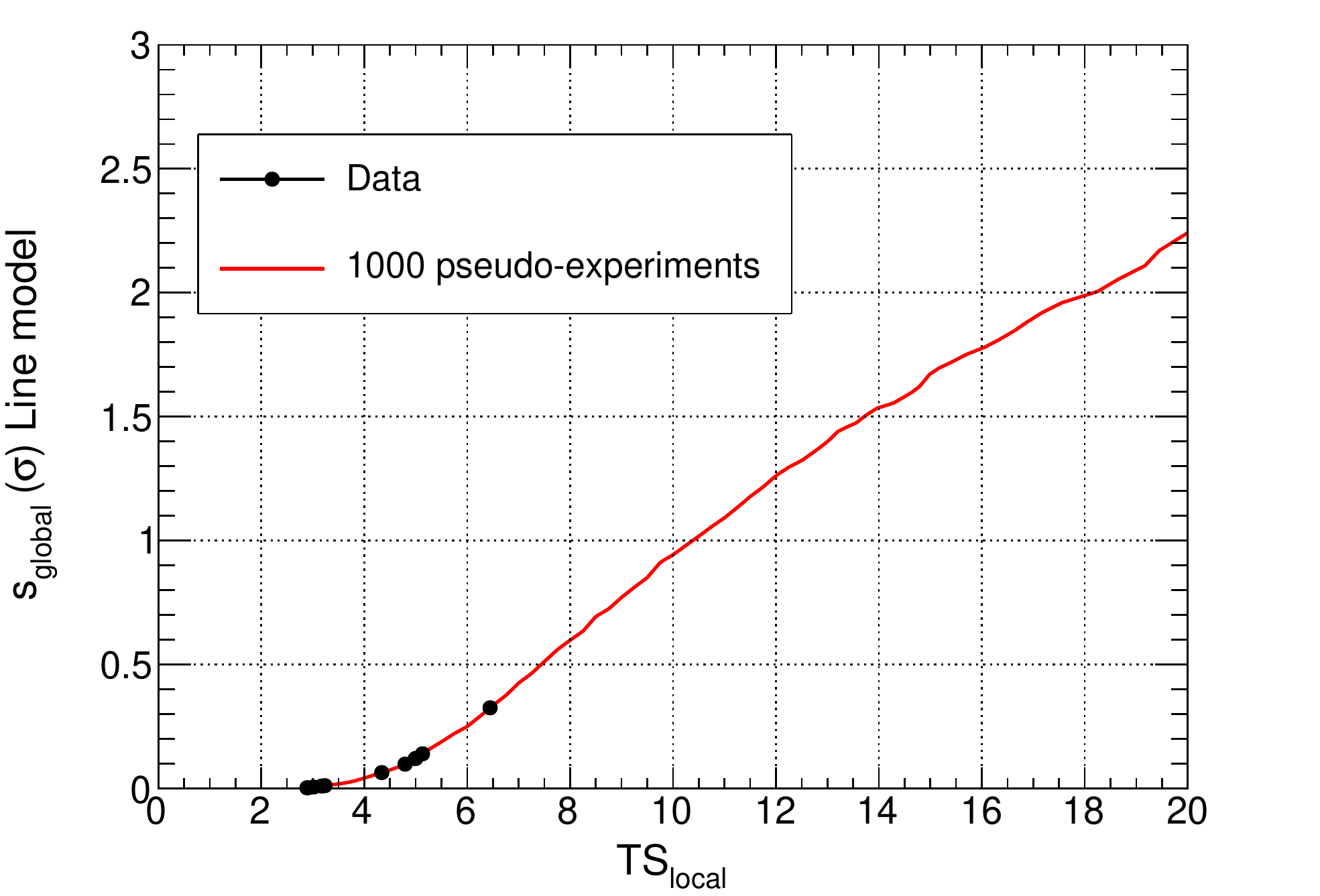} &
\includegraphics[width=0.32\textwidth,height=0.18\textheight,clip]{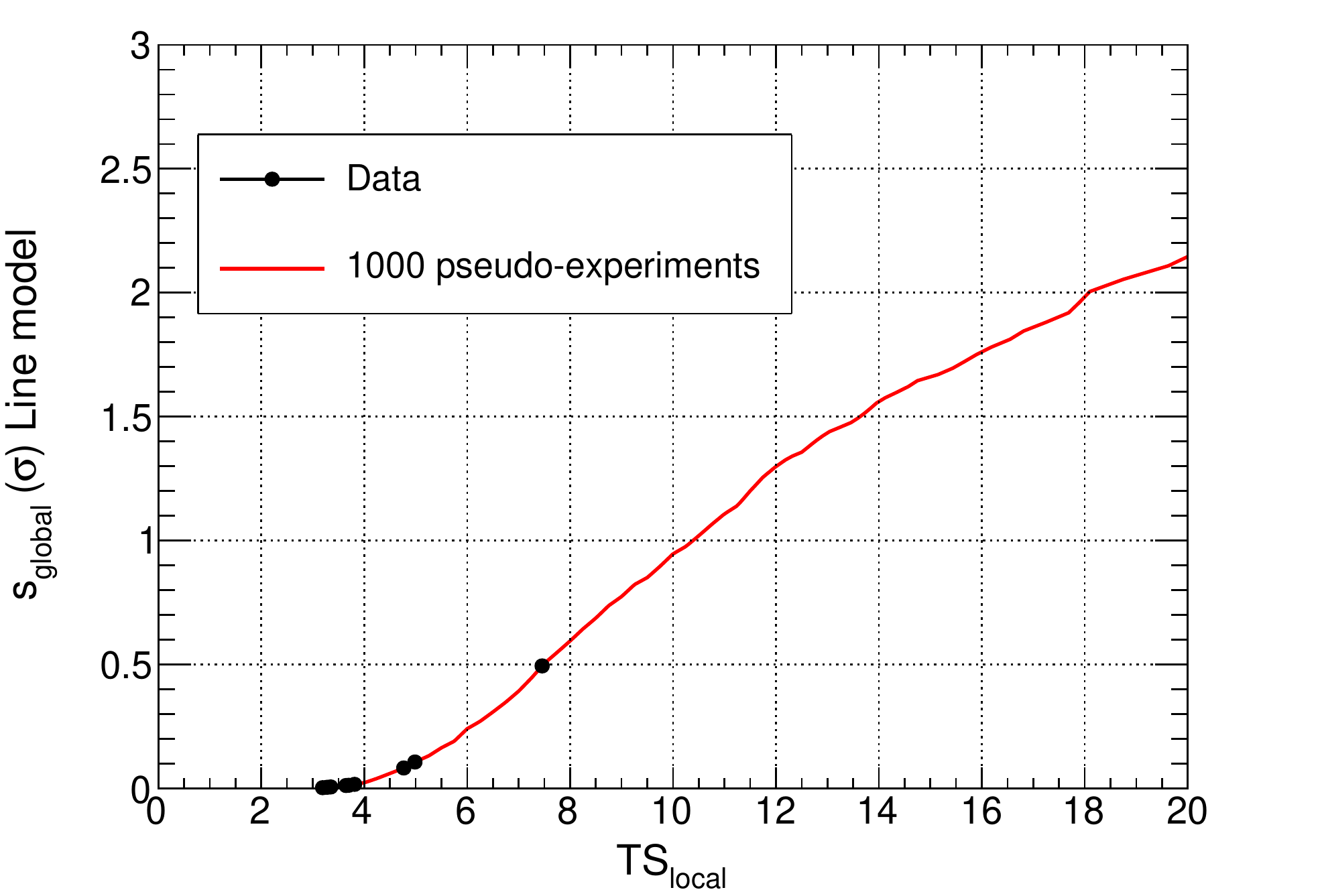} \\
\includegraphics[width=0.32\textwidth,height=0.18\textheight,clip]{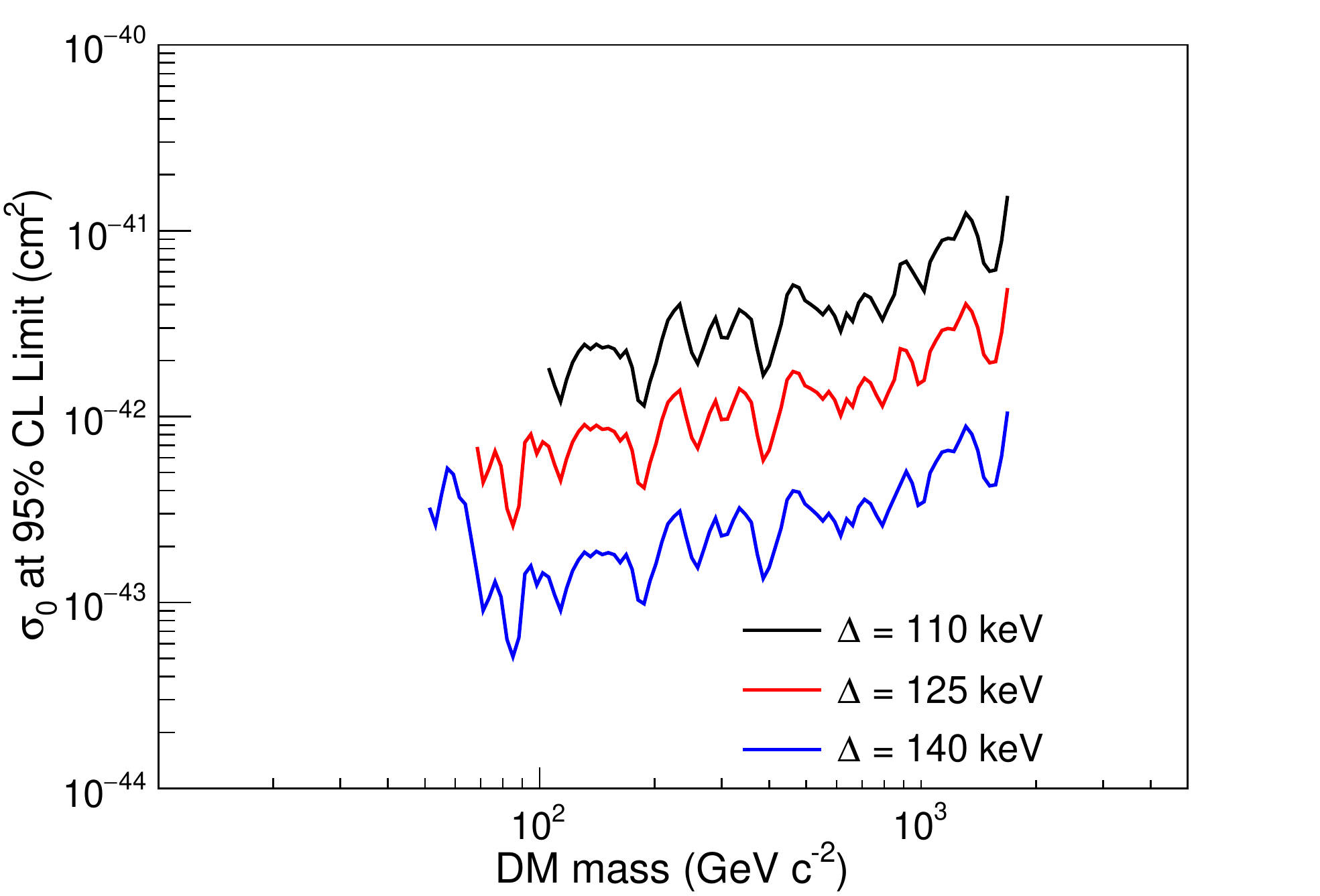} &
\includegraphics[width=0.32\textwidth,height=0.18\textheight,clip]{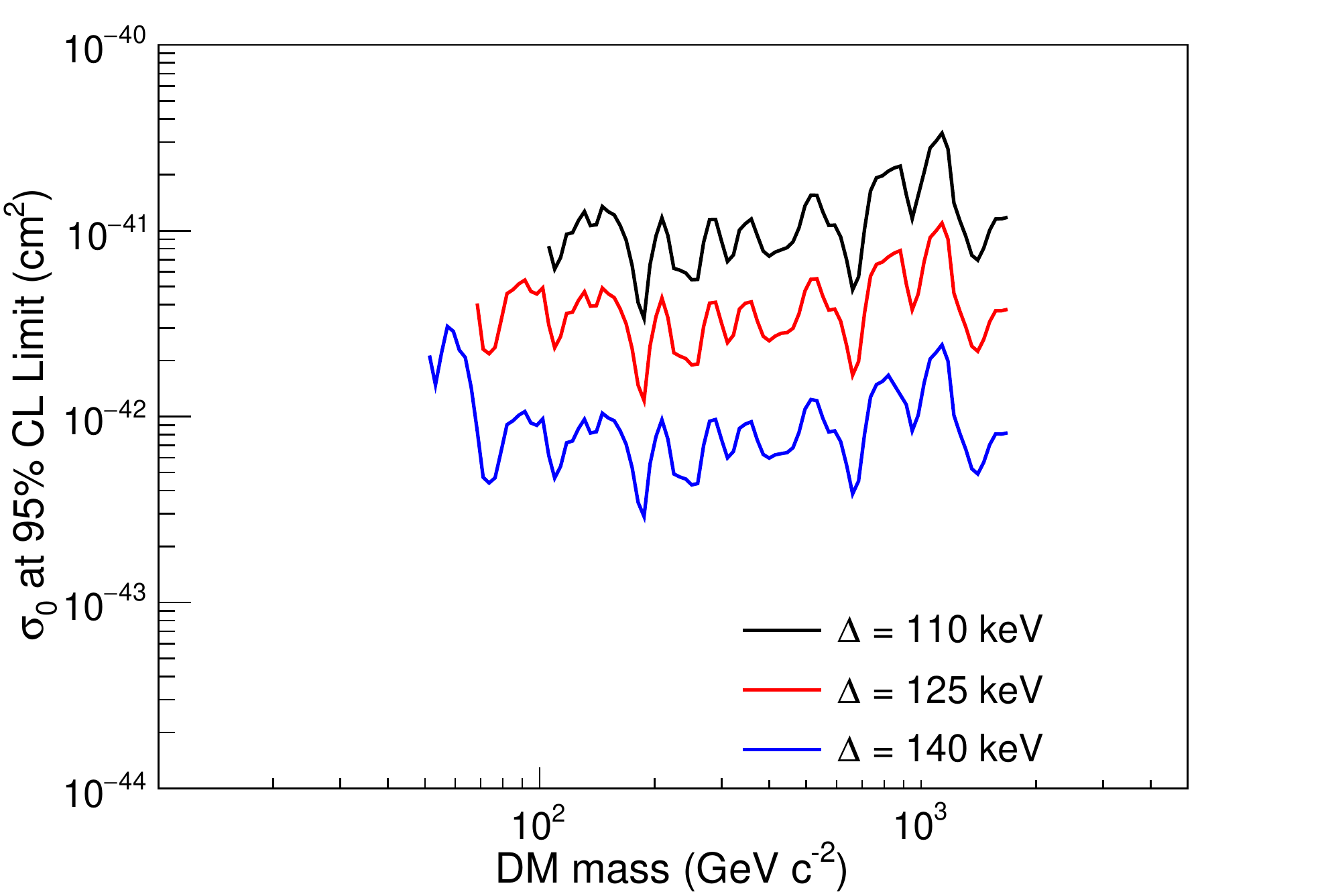} &
\includegraphics[width=0.32\textwidth,height=0.18\textheight,clip]{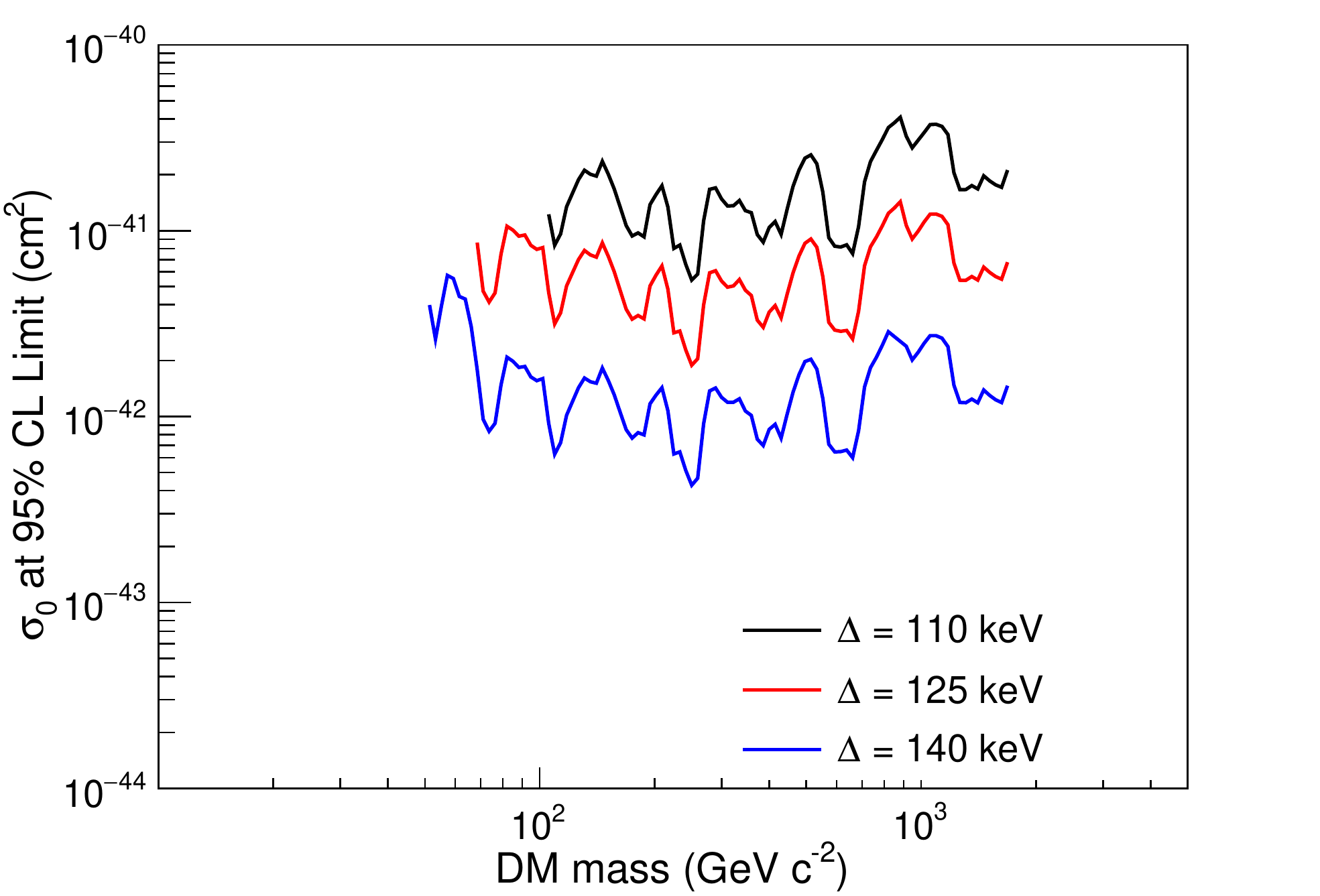}\\
\end{tabular}
\caption{Results of the search for line-like features in $10\degrees$ (left column), $30\degrees$ (middle column) and $45\degrees$ (right column) RoIs. The first row shows the upper limit at $95\%$ CL on the line intensity compared with the expectations bands from the pseudo-experiments (the green and yellow bands show the central $68\%$ and $95\%$ expectation bands for the $95\%$ CL limit). The second row shows the value of the local TS in the various energy windows compared with the expectations from the pseudo-experiments (the green and yellow bands show the one-sided $68\%$ and $95\%$ expectation bands for the TS local). The third row shows the conversion from the local TS to the global significance. The last row shows the upper limits on the cross section per nucleon $\sigma_0$ for DM annihilation to $e^+e^-$ via inelastic scattering as a function of energy, calculated for three values of the splitting mass parameter, i.e. $\Delta=110, 125$ and $140\units{keV}$. }
\label{fig:line}
\end{figure*}

\medskip

Figure~\ref{fig:line} shows the results obtained with the line model for the features (i.e. the inelastic scattering scenario). The results for $10\degrees$, $30\degrees$ and $45\degrees$ RoIs are summarized in the left, middle and right column respectively. The upper limits (ULs) of the intensity of the line-like feature at $95\%$ CL are shown in the first row, where they are compared with the confidence belts evaluated using the pseudo-experiment technique described in Sec.~\ref{sec:anamet}. In most cases the limits lie within the central $95\%$ confidence belt. The second row of Figure~\ref{fig:line} shows the local significance of possible features, while the third row shows the conversion from the local test statistics $TS_{\text{local}}$ to the global significance $s_{\text{global}}$. Also in this case, the most significant features have global significances less than $1.5 \sigma$.

Starting from the upper limits of the first row and using eq.~\ref{eq:idm} with the assumption that the capture rate in eq.~\ref{eq:eq5} scales linearly with the DM-nucleon cross section, we evaluate the limits on the DM-nucleon cross-section as:

\begin{equation}
    \sigma_{\text{0,UL}}(m_\chi) = \frac{I_{\text{UL}}(E=m_\chi)~\Delta\Omega}{\Phi_{\text{DM}}(E=m_\chi)} \times 10^{-40} \units{cm^2}
\end{equation}
where $\Delta\Omega$ is again the solid angle corresponding to the selected RoI, $\Phi_{\text{DM}}(E)$ is taken from Eq.~\ref{eq:idm} divided by $2$ and $10^{-40} \units{cm^2}$ is the DM-nucleon cross section value used to calculate the capture rate shown in Fig.~2 of Ref.~\cite{Menon:2009qj}. 

The last row in Figure~\ref{fig:line} shows the limits on the cross section per nucleon $\sigma_0$ for DM annihilation to $e^+e^-$ via inelastic scattering as a function of DM mass. These limits have been calculated for three  values of the splitting mass parameter, $\Delta=110,~125$ and $140\units{keV}$. The limits on $\sigma_0$ range from about $10^{-43}\units{cm^2}$ to $10^{-41}\units{cm^2}$, and they are slightly lower than those calculated in our previous work~\cite{Ajello:2011dq}. 

We point out here that in Ref.~\cite{Ajello:2011dq} a different analysis technique was used with respect to the present work, and systematic errors were not included in the analysis. The inclusions of systematic errors has a significant impact on the upper limits, which are of at least one order of magnitude worse than those obtained when systematic errors are neglected. Thus, since  the error is dominated by the systematic component, the larger data set used here did not result in stronger limits with respect to~\cite{Ajello:2011dq}, in particular in the low energy range explored in the  current analysis. On the other hand, with this larger data set we are able to better constrain the DM-nucleon cross section for higher DM masses.

\section{Conclusion}
\label{sec:con}
In this work we have derived upper limits on the DM-nucleon scattering cross section using the Fermi-LAT data on CREs from the Sun. We have analyzed the data within the framework of two models, in which CREs are produced from the annihilations of DM particles captured in the Sun either directly or indirectly, via a long-lived mediator. 

In case of the long-lived mediator scenario, the limits on the spin-independent cross-section are in the range from about $10^{-46}\units{cm^2}$ to about $10^{-44}\units{cm^2}$, while those on the spin-dependent are in the range from about $10^{-43}\units{cm^2}$ to about $10^{-41}\units{cm^2}$ for DM masses between $48\units{GeV}$ and $1.7\units{TeV}$ and RoIs larger than $10\degrees$. The limits depend on the decay length of the mediator and change by a factor 2 for values of the decay length between $R_{\odot}$ and $5 \units{AU}$. 
For what concerns the other parameters in the model, the limits are expected to scale linearly with the local DM density $\rho_{\odot}$ and to exhibit a mild dependence on $v_{\odot}$ and $v_{rms}$.

\begin{figure}[!t]
\centering
\includegraphics[width=0.99\columnwidth,height=0.25\textheight,clip]{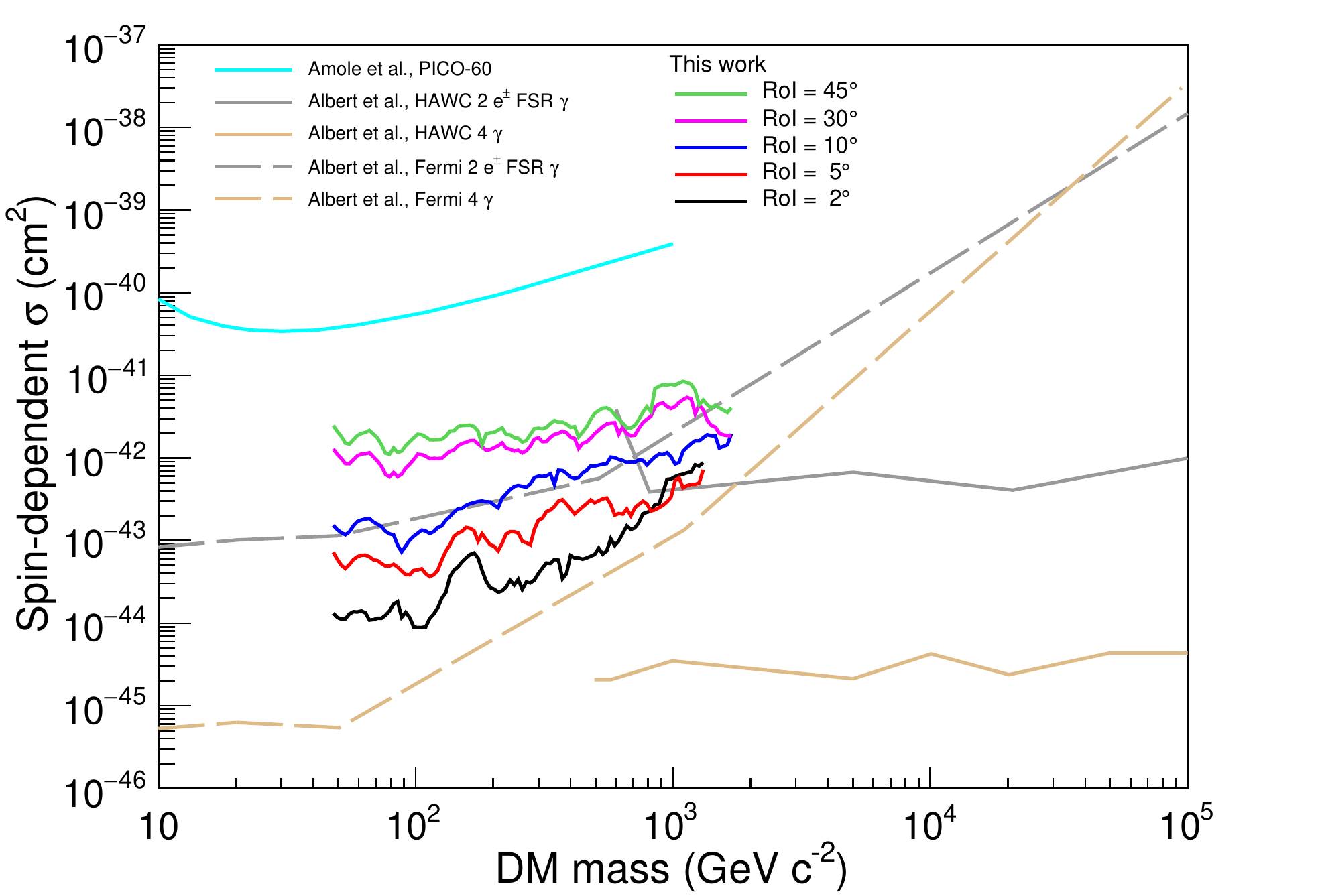}
\caption{Upper limits at 95\% CL on the spin-dependent DM-nucleon scattering cross section for the long lived mediator with decay length $L=R_{\odot}$. The current results with different RoIs, from $2$ to $45\degrees$, are shown together with those from the HAWC and Fermi~\cite{Albert:2018jwh} with gamma rays from DM annihilating into 2 $e^{\pm}$, i.e., 2 $e^{\pm}$ FSR $\gamma$ case, (gray lines) and into 4 $\gamma$ (brown lines). The cyan line shows the PICO-60 limits at $90\%$ CL~\cite{Amole:2019fdf}.} 
\label{fig:expcomp}
\end{figure}

A summary of the limits on the spin-dependent DM-nucleon cross section for the long-lived mediator scenario with decay length $L=R_{\odot}$ is shown in Fig.~\ref{fig:expcomp}, together with the recent HAWC and Fermi~\cite{Albert:2018jwh} results obtained with gamma rays constraints, i.e., for the two cases
in which the mediator decays into 2 $\gamma$ (4$\gamma$ case), and the case as the one discussed in this work of mediator decaying into $e^{\pm}$ and 
where $\gamma$s are produced via FSR (i.e., 2 $e^{\pm}$ FSR $\gamma$ case).
The plot also shows the PICO-60 limit at $90\%$ CL~\cite{Amole:2019fdf}.
In Fig.~\ref{fig:expcomp} we show the limits obtained from the present analysis using different RoIs, from $2\degrees$ to $45\degrees$. The smaller RoIs yield tighter constraints, but it should be kept in mind that CREs suffer the effects of the interplanetary magnetic field. However, even with a conservative choice of the RoI, the limits obtained from our analysis are competitive with those obtained from the solar gamma rays by Fermi and HAWC assuming DM annihilations in $2 e^{\pm}$ for DM masses in the$\units{TeV}$ range. 

As mentioned in section~\ref{sec:inter}, our results have been derived assuming equilibrium between DM capture and annihilation in the Sun. If the equilibrium is not reached, the annihilation rate is reduced of a factor $\tanh^2(t/\tau)$, and consequently the limits on the DM-nucleon cross sections must be scaled of a factor $1/\tanh^2(t/\tau)$. In Refs.~\cite{Jungman:1995df, Griest:1986yu} it is shown that  
$t/\tau \propto \langle \sigma_{ann} v \rangle^{1/2} \times \sigma^{1/2}$, where $\langle \sigma_{ann}v \rangle$ is the velocity-averaged DM annihilation cross section and $\sigma$ is the DM-nucleon scattering cross section. Assuming for the lifetime of the Sun the value $t=4.5\units{Gyr}$ and for the velocity-averaged cross section the value $\langle \sigma_{ann} v \rangle =3 \times 10^{-26} \units{cm^3/s}$ (thermal value), and rescaling the capture rates shown in Figure~\ref{fig:fig2}, we have evaluated the minimum values of $\sigma$ needed to attain equilibrium between capture and annihilation at the present time. 
These values are  $\sim 10^{-45}\units{cm^2}$ for the spin-independent case and $\sim 10^{-43}\units{cm^2}$ for the spin-dependent case in the DM mass range explored in the present work, and are of the same order of magnitude as the limits that we set using the larger RoIs. 

On the other hand, the limits that we set for smaller RoIs (i.e. $2\degrees$ and  $5\degrees$ in Fig.~\ref{fig:expcomp}) and for low DM masses should be corrected to take non-equilibrium into account. We find that, assuming $\langle \sigma_{ann} v \rangle =3 \times 10^{-26} \units{cm^3/s}$, these limits should be rescaled by a factor of few. If $\langle \sigma_{ann} v \rangle$ is lower than the thermal value, the limits will change accordingly. However, the current limits on $\langle \sigma_{ann} v \rangle$ obtained from different analyses (see for instance~\cite{Ackermann:2013yva,Ackermann:2015zua,Mazziotta:2017ruy,Zu:2017dzm}) are of the same order of the thermal value for DM masses up to about $100 \units{GeV}$ and are larger for higher DM masses.

Also, a further point to mention is that the limits in the elastic scattering scenario are evaluated assuming that DM particles annihilate only into a pair of light mediators $\phi$s, which in turn decay into $e^+ + e^-$ pairs, while those in the inelastic scattering scenario are evaluated assuming that DM particles annihilate only into $e^+ + e^-$ pairs. If annihilation into $e^+ e^-$ is only a fraction, with  branching ratio $BR$, of the total annihilation cross-section,  then the constraints on the scattering cross sections will rescale as $1/BR$.

In case of the inelastic scattering scenario, the limits on the DM-nucleon cross-section lie in the range from about $10^{-43}\units{cm^2}$ to about $10^{-41}\units{cm^2}$ for the splitting mass parameter in the range $110-140\units{keV}$, with the capture rates taken from Ref.~\cite{Menon:2009qj}. These limits are consistent or even stronger than those derived by the CDMS experiment assuming a mass splitting parameter of $120\units{keV}$~\cite{Ahmed:2010hw}. Finally, we point out that this analysis would benefit from a dedicated calculation of the capture rates for the full DM mass range explored here and different values of the mass splitting parameter.

\medskip

\begin{acknowledgments} 
The Fermi LAT Collaboration acknowledges generous ongoing support
from a number of agencies and institutes that have supported both the
development and the operation of the LAT as well as scientific data analysis.
These include the National Aeronautics and Space Administration and the
Department of Energy in the United States, the Commissariat \`a l'Energie Atomique
and the Centre National de la Recherche Scientifique / Institut National de Physique
Nucl\'eaire et de Physique des Particules in France, the Agenzia Spaziale Italiana
and the Istituto Nazionale di Fisica Nucleare in Italy, the Ministry of Education,
Culture, Sports, Science and Technology (MEXT), High Energy Accelerator Research
Organization (KEK) and Japan Aerospace Exploration Agency (JAXA) in Japan, and
the K.~A.~Wallenberg Foundation, the Swedish Research Council and the
Swedish National Space Board in Sweden.
 
Additional support for science analysis during the operations phase is gratefully
acknowledged from the Istituto Nazionale di Astrofisica in Italy and the Centre
National d'\'Etudes Spatiales in France. This work performed in part under DOE
Contract DE-AC02-76SF00515.
\end{acknowledgments}

\bibliographystyle{apsrev4-1}
\bibliography{LATCRESun.bib}{}

\end{document}